\documentclass[sigplan,10pt]{acmart}

\usepackage{booktabs}
\usepackage{subcaption}

\let\terms\undefined
\usepackage[utf8]{inputenc}
\usepackage[T1]{fontenc}
\usepackage{lmodern}
\usepackage{wrapfig}
\usepackage{amsmath}
\usepackage{amssymb}
\usepackage{array}
\usepackage{multirow}
\usepackage{multicol}
\usepackage{todonotes}
\usepackage{booktabs}
\usepackage{pifont}
\usepackage{tikz}
\usepackage{color}
\usepackage{float}
\usepackage{colortbl}
\usepackage{ltl}
\usepackage{xspace}
\usepackage{bussproofs}
\usepackage[capitalise,nameinlink]{cleveref}
\usepackage{varwidth}
\usepackage{stmaryrd}
\usepackage{etex}
\usepackage{listings}
\usepackage{amsmath}

\DeclareMathDelimiter{(}{\mathopen} {operators}{"28}{largesymbols}{"00}
\DeclareMathDelimiter{)}{\mathclose}{operators}{"29}{largesymbols}{"01}

\usetikzlibrary{calc}
\usetikzlibrary{automata}
\usetikzlibrary{backgrounds}
\usetikzlibrary{decorations.pathreplacing}
\usetikzlibrary{shapes,arrows}
\usetikzlibrary{positioning}
\usetikzlibrary{shadows}

\newcommand{\logic}{\ensuremath{\mathbb{L}}\xspace}
\newcommand{\mutex}{\ensuremath{\mathbb{U}}\xspace}
\newcommand{\nats}{\ensuremath{\mathbb{N}}}
\newcommand{\signal}{\ensuremath{s}}
\newcommand{\dtime}{\mbox{\ensuremath{\mathcal{T}\hspace{-2.5pt}\scalebox{0.8}{\ensuremath{\boldsymbol{i}\boldsymbol{m}\boldsymbol{e}}}}}}
\newcommand{\from}{\ensuremath{\colon}}

\renewcommand{\to}{\ensuremath{\rightarrow}}
\newcommand{\values}{\ensuremath{\mathcal{V}}}
\newcommand{\inputs}{\ensuremath{\mathcal{I}}}
\newcommand{\outputs}{\ensuremath{\mathcal{O}}}
\newcommand{\signals}{\ensuremath{\mathcal{S}}}
\newcommand{\cells}{\ensuremath{\mathbb{C}}\xspace}
\newcommand{\bool}{\ensuremath{\mathcal{B}}}

\newcommand{\cfm}{\ensuremath{\mathcal{M}}\xspace}
\newcommand{\branch}[2]{\ensuremath{#1 \! \wr \hspace{-0.6pt} #2}}
\newcommand{\term}{\ensuremath{\tau}}
\newcommand{\fterm}{\ensuremath{\term_{F}}\xspace}
\newcommand{\pterm}{\ensuremath{\term_{P}}\xspace}

\newcommand{\name}[1]{{\text{\texttt{#1}}}}

\newcommand{\terms}{\ensuremath{\mathcal{T}}\xspace}

\newcommand{\sep}{\ensuremath{\quad | \quad}}
\newcommand{\upd}[2]{\ensuremath{\llbracket \hspace{1pt} #1 \leftarrowtail \, #2 \hspace{1pt} \rrbracket}}
\newcommand{\sats}{\ensuremath{\;\vDash_{\!\langle \hspace{-1pt} \cdot \hspace{-1pt} \rangle}}}

\newcommand{\set}[1]{\ensuremath{\{ #1 \}}}

\newcommand{\pterms}{\ensuremath{\terms_{\!P\hspace{-0.5pt}}}\xspace}
\newcommand{\fterms}{\ensuremath{\terms_{\!F\hspace{-0.5pt}}}\xspace}

\newcommand{\functions}{\ensuremath{\mathcal{F}}\xspace}
\newcommand{\predicates}{\ensuremath{\mathcal{P}}\xspace}
\newcommand{\assign}[1]{\ensuremath{\langle #1 \rangle}}

\newcommand{\TSL}{\text{TSL}\xspace}

\newcommand{\inames}{\ensuremath{\mathbb{I}}\xspace}
\newcommand{\onames}{\ensuremath{\mathbb{O}}\xspace}
\newcommand{\pnames}{\ensuremath{\mathbb{P}}\xspace}
\newcommand{\fnames}{\ensuremath{\mathbb{F}}\xspace}

\newcommand{\vertices}{\ensuremath{V}}
\newcommand{\labeling}{\ensuremath{\ell}}
\newcommand{\dependencies}{\ensuremath{\delta}}

\newcommand{\arrComp}{\ensuremath{>\mkern-9mu>\mkern-9mu>} }

\newcommand{\codeinline}[1]{\lstinline{#1}}

\newcommand{\eg}{{\em e.g.~\xspace}}

\lstdefinestyle{tsl}{
  language=c,
  basicstyle=\footnotesize\ttfamily,breaklines=true,
  mathescape=false,
  frame=none,
  literate=
    {->}{{{\color{blue!80}\ \ \texttt{->}\;\ }}}1
    {<-}{{{\color{red!50!black}\ \ \texttt{<-}\;\ }}}1
    {<->}{{{\color{blue!80}\ \ \texttt{<->}\;\ }}}1
    {\&\&}{{{\color{blue!80}\ \ \texttt{\&\&}\;\ }}}1
    {||}{{{\color{blue!80}\ \ \texttt{||}\;\ }}}1
    {!}{{{\color{blue!80}\ \ \texttt{!}\;\ }}}1
    {[}{{{\color{red!50!black}\ \ \texttt{[}\;\ }}}1
    {]}{{{\color{red!50!black}\ \ \texttt{]}\;\ }}}1
    {\{}{{{\color{black}\ \ \texttt{\{}\;\ }}}1
    {\}}{{{\color{black}\ \ \texttt{\}}\;\ }}}1
    {()}{{\ \ \texttt{()}\;\ }}1
    {;}{{\ \ \texttt{;}\;\ }}1,    
  deletekeywords={Double,Int,init,Bool},
  identifierstyle={\ttfamily\color{black}},
  keywordstyle=[1]{\ttfamily\color{green!60!black}},
  keywordstyle=[2]{\ttfamily\color{violet}},
  keywordstyle=[3]{\ttfamily\color{blue!80}},
  keywordstyle=[4]{\ttfamily\color{yellow!50!black}},
  keywordstyle=[5]{\ttfamily\color{red}},
  keywordstyle=[6]{\ttfamily\color{cyan!40!black}},
  morekeywords=[1]{COUNTUP, COUNTDOWN, INCMIN, INCSEC, IDLE, ZERO, RESET, COUNTING, ANYKEY, START, STARTANDMIN, STARTANDSEC},
  morekeywords=[2]{xor, press, tillAnyInput},
  morekeywords=[3]{X, W},
  morekeywords=[4]{x, y},
  morekeywords=[6]{always, guarantee, initially},
  commentstyle=\color{orange!50!black}, 
  stringstyle=\color{red!50!black},
}

\setcopyright{none}

\makeatletter\if@ACM@journal\makeatother
\acmJournal{PACM}
\acmVolume{1}
\acmNumber{1}
\acmArticle{1}
\acmYear{2017}
\acmMonth{1}
\acmDOI{10.1145/nnnnnnn.nnnnnnn}
\startPage{1}
\else\makeatother
\acmConference[]{ACM SIGPLAN Conference}{2019}{}
\acmYear{2019}
\acmISBN{978-x-xxxx-xxxx-x/YY/MM}
\acmDOI{10.1145/nnnnnnn.nnnnnnn}
\startPage{1}
\fi

\begin{document}

\title[]{Synthesizing Functional Reactive Programs}

\author{Bernd Finkbeiner}
\orcid{1234-5678-9012}
\affiliation{%
  \institution{Saarland University}
  \state{Germany} 
}

\author{Felix Klein}
\orcid{1234-5678-9012}
\affiliation{%
  \institution{Saarland University}
  \state{Germany} 
}

\author{Ruzica Piskac}
\orcid{1234-5678-9012}
\affiliation{%
  \institution{Yale University}
  \state{CT, USA} 
}

\author{Mark Santolucito}
\orcid{1234-5678-9012}
\affiliation{%
  \institution{Yale University}
  \state{CT, USA} 
}

\begin{abstract}
Functional Reactive Programming (FRP) is a paradigm that has simplified the construction of reactive programs.
There are many libraries that implement incarnations of FRP, using abstractions such as Applicative, Monads, and Arrows. However, finding a good control flow, that correctly manages state and switches behaviors at the right times, still poses a major challenge to developers.

An attractive alternative is specifying the behavior instead of programming it, as made possible by the recently developed logic: Temporal Stream Logic (TSL).
However, it has not been explored so far how Control Flow Models (CFMs), as synthesized from TSL specifications, can be turned into executable code that is compatible with libraries building on FRP. We bridge this gap, by showing that CFMs are indeed a suitable formalism to be turned into Applicative, Monadic, and Arrowized FRP.

We demonstrate the effectiveness of our translations on a real-world kitchen timer application, which we translate to a desktop application using the Arrowized FRP library \texttt{Yampa}, a web application using the Monadic \texttt{threepenny-gui} library, and to hardware using the Applicative hardware description language \texttt{ClaSH}.
\end{abstract}

%\keywords{Reactive Synthesis, FRP}

\maketitle

\section{Introduction}
\label{sec:intro}

Reactive programs implement a broad class of computer systems whose defining element is  the continued interaction between the system and
its environment. Their importance can be seen through the wide range
of applications, such as embedded devices~\cite{Helbling2016}, games~\cite{Perez17}, robotics~\cite{JingTYK16}, hardware circuits~\cite{Khalimov2014ParameterizedSynthesisCaseStudyAMBA}, GUIs~\cite{elm}, and interactive multimedia~\cite{Santolucito15ICMC}.

Functional Reactive Programming (FRP) is a paradigm for writing programs for reactive systems.
The fundamental idea of FRP is to extend the classic building blocks of functional programming
with the abstraction of a $signal$ to describe values varying over time.
In contrast to sequential programs being executed step by step, FRP programs lead to stream processing networks that manage state and switch between behaviors dependent on the user input.
Therefore, FRP programs can be exceptionally efficient.
For example, a network controller recently implemented as an FRP program on a multicore processor outperformed all its contemporary competing implementations~\cite{voellmy2013maple}.

Building a reactive program is a complex process, of which the most difficult part is finding a good and correct high-level design~\cite{PitermanPS06}.
Furthermore, even once this design has been fixed, implementing the system still remains a highly error-prone process~\cite{Shan16}.
While FRP helps with the latter problem by using an advanced type system to introduce a clear concept of time, it leaves the challenge of switching between behaviors and managing state efficiently open.
The use of temporal logic has been explored to test properties of FRP programs~\cite{DBLP:journals/pacmpl/PerezN17}, however testing still leaves space for possible errors.

Another solution for solving the design challenge has been proposed with Temporal Stream Logic~\cite{tsl}, a specification logic to specify the temporal control flow behavior of a program. The logic enforces a clean separation between control and data transformations, which also can be leveraged in FRP~\cite{hudakFRAN}. Temporal Stream Logic (TSL) is used in combination with a reactive synthesis engine to automatically create an abstract model of temporal control called a Control Flow Model (CFM), which satisfies the given specification. TSL combines Boolean and temporal operations with \textit{predicates}~$ \name{p}~\name{s}_{\name{i}} $, evaluated on input signals~$ \name{s}_{\name{i}} $, and \textit{updates} denoted by~$ \upd{\name{s}_{\name{o}}}{\name{f}~\name{s}_{\name{i}}} $, which map pure functions~$ \name{f} $ to input signals~$ \name{s}_{\name{i}} $ and pipe the result to a signal~$ \name{s}_{\name{o}} $. An example for a TSL specification is given by
\begin{equation*}
  \begin{array}{l}
    \LTLglobally \big(\, \text{(}\name{event}~\name{click} \ \leftrightarrow \ \upd{\name{count}}{\name{increment}~\name{count}} \text{)} \\ \qquad \wedge \ \upd{\name{screen}}{\name{display}~\name{count}} \, \big)
  \end{array}
\end{equation*}
which states that a counter must be incremented whenever there is a
click event, while the value of the counter is displayed on a
screen.

\begin{figure}[t]
\centering
\begin{minipage}{0.6\textwidth}
  \begin{lstlisting}
yampaButton :: SF (Event MouseClick) Picture
yampaButton = proc click -> do
  rec
    count    <- init 0 -< newCount
    newCount <- arr f1 -< (click, count)
    pic      <- arr f2 -< count
  returnA -< pic

f1 :: (Event MouseClick, Int) -> Int
f1 (click, count)
  | isEvent click = count + 1
  | otherwise     = count

f2 :: Int -> Picture
f2 count = render count
  \end{lstlisting}
\end{minipage}
\vspace{-1em}
\caption{A button written with the FRP library \texttt{Yampa}.}
\label{fig:button}
\end{figure}

An implementation that satisfies the specified behavior, built using the FRP library \texttt{Yampa}~\cite{courtney2003yampa}, is shown in \cref{fig:button}.
The program implements a button in a GUI which shows a counter value that increments with every click.
The \texttt{Yampa} FRP library uses an abstraction called arrows, where the arrows define the structure and connection between functions~\cite{liu2007plugging}.
As mentioned before, they can be used to cleanly separate data transformations into pure functions, creating
a visually clear separation between the control flow and the data level.
In the example program in Fig.~\ref{fig:button}, this separation is clearly visible.
The top half is the ``arrow'' part of the code, which defines a control flow.
The bottom half is the ``functional'' part of the code, which defines the functions {\tt{f1}} and {\tt{f2}}, describing pure data transformations.

In the TSL specification, function applications, like \name{click},
\name{increment} or \name{display}, are not tied to a particular
implementation. Instead, the synthesis engine ensures that the
specification is satisfied for all possible implementations that may
be bound to these placeholders, similar to an unknown polymorphic
function that is used as an argument in a functional program. Thus,
the implementation of \cref{fig:button} indeed satisfies the given
specification by implementing \name{event} with \name{isEvent},
\name{increment} with \name{(+1)}, and \name{display} with
\name{render}.

The immediate advantage of synthesis over manual programming is that if the synthesis succeeds, there is a guarantee that the constructed program satisfies the specification. Sometimes, the synthesis does not succeed, and this also leads to interesting results. An example is given by the FRPZoo~\cite{FRPzoo} study, which consists of implementations for the same program for 16 different FRP libraries. The program consists of
two buttons that can be clicked on by the user:
a \name{counter} button, which keeps track of the number of clicks,
and a \name{toggle} button, which turns the counter button on and off.
To our surprise, after translating the written-English specification from the FRPZoo website into a formal \TSL specification,
the synthesis procedure was not able to synthesize a satisfying program.
By inspecting the output of the synthesis tool, we noticed that the specification is actually unrealizable.
The problem is that the specification requires
the counter to be incremented whenever the \texttt{counter} button is clicked
and, to be reset to zero whenever the \texttt{toggle}
button is clicked. This creates a conflict when both
buttons are clicked at the same time. To obtain a solution, we had to
add the assumption that both buttons are never pressed simultaneously.

While the work of Finkbeiner et al.\, 2019 discusses the synthesis process
for creating the CFM in detail, it does not elaborate on how a CFM is actually turned into FRP code.
In this work we explore this process, and show how Causal Commutative Arrows (CCA) form a foundational abstraction for FRP in the context of program synthesis.
From this connection between CCA and a CFM, we then build a system to generate library-independent FRP code across a range of FRP abstractions.

There is no single style of FRP which is generally accepted as canonical.
Instead, FRP is realized through a number of libraries, which are based on fundamentally different abstractions, such as Monadic FRP~\cite{ploeg2015frpnow,threepennygui, Reflex}, Arrowized FRP~\cite{courtney2003yampa,perez2016yampa,UISF}, and Applicative FRP~\cite{clash2015,reactivebanana}.
We show how our system is flexible enough to handle these various abstractions by demonstrating the translation from a CFM to \texttt{Yampa}~\cite{courtney2003yampa,perez2016yampa}, \texttt{threepenny-gui}~\cite{threepennygui}, and \texttt{ClaSH}~\cite{clash2015}.
We do not imagine FRP synthesis as a replacement for FRP libraries, but rather as a complement.
Through synthesis and code generation, users can automatically construct FRP programs in these libraries, which provide critical interfaces to input/output domains.
Furthermore, we show that $\TSL$ synthesis generates code as expressive as CCA.
While this power is sufficient for many applications, the FRP libraries still provide an interface to more powerful language abstraction features, in cases the user chooses to use them.

\medskip

\noindent In summary, the paper makes the following contributions:

\begin{enumerate}
\item We describe the process of automatically generating library independent FRP control code from TSL specifications.

\item We examine the relation between CFMs and CCA, and compare the differences between various FRP abstractions during the translation process.

\item We demonstrate our translations on a real-world kitchen timer application, build as a desktop application using the Arrowized FRP library \texttt{Yampa}, as a web application using the Monadic library \texttt{threepenny-gui}, and to hardware using the Applicative hardware description language \texttt{ClaSH}.

\item We provide an open-source tool for the synthesis of FRP programs from TSL specification\footnote{\url{https://github.com/reactive-systems/tsltools}}

\end{enumerate}

\goodbreak

\section{Preliminaries}

We assume time to be discrete and denote it by the set $ \dtime $ of
positive integers. A value is an arbitrary object of arbitrary type,
where we use $ \values $ to denote the set of all values. We consider
the Boolean values $ \bool \subseteq \values $ as a special subset,
which are either $ \name{true} \in \bool $ or
$ \name{false} \in \bool $.

A signal~$ \signal \from \dtime \to \values $ is a function fixing a
value at each point in time. The set of all signals is denoted
by~$ \signals $, usually partitioned into input signals $ \inputs $
and output signals $ \outputs $.

An $ n $-ary function~$ f \from \values^{n} \to \values $ determines a
new value from $ n $ given values. We denote the set of all functions
(of arbitrary arity) by~$ \functions $.  Constants are  functions of
arity zero, while every constant is also a value, i.e, an element of
$ \functions \cap \values $. An $ n $-ary
predicate~$ p \from \values^{n} \to \bool $ checks a truth statement
on $ n $ given values.  The set of all predicates (of
arbitrary arity) is denoted by~$ \predicates $.

\subsection{Functional Reactive Programming}
\label{sec:frp}

FRP is a programming paradigm that uses the abstractions available in functional programming to create an abstraction of time.
The core abstraction in FRP is that of a signal
\begin{equation*}
 Signal\ a :: \dtime \to a
\end{equation*}
which produces values of some arbitrary type $ a $ over time. The
type $ a $ can be an input from the world, such as the current
position of the mouse, or an output type, such as some text that
should be rendered to the screen. Signals are also used internally to
manipulate values over time, for example if the position of the
mouse should be rendered to the screen.

\paragraph{Arrows}
There are many incarnations of FRP, which use various abstraction to manipulate signals over time.
One popular abstraction for FRP is a Monad, but a weaker abstraction called Arrows, is also used in many modern libraries~\cite{perez2016yampa, murphy2016livefrp}.
The Arrow abstraction describes a computation connecting inputs and outputs in a single type~\cite{hughes2000generalising}.
Hence, an Arrow type also allows us to describe reactive programs that process inputs and produce outputs over time.

\begin{figure}
\centering
\scalebox{1}{\begin{tikzpicture}[%
    >=stealth' % here you can choose the arrow tip of your choice
		]
\tikzset{% define styles for the different boxes, adjust parameters to wish
  base/.style={align=center, minimum height=5mm, minimum width=10mm, draw=black,thick},
	circ/.style={align=center, minimum width=4mm, circle, draw=black,thick},
	% coord node style is used for drawing the arrows i.e. invisible nodes
  coord/.style={coordinate},
}
	
\matrix [column sep=10mm, row sep=14mm, every node/.style={
    shape=rectangle,
    text width=2.75cm,
    text centered,
    very thick,
    draw=none,
}] {
\node (a) {(a) arr f}; &
\node (b) {(b) a1 \arrComp a2};\\
\node (c) {(c) first a}; &
\node (d) {(d) loop a};\\
};

\node [base,minimum width=12mm,minimum height=9mm,above=0mm of a,draw=black,dotted,thin](abox) {f};
\node [base,above left=2mm and 3mm of b.north](bbox1) {a1};
\node [base,above right=2mm and 3mm of b.north](bbox2) {a2};
\node [base,above=4mm of c](cbox) {a};
\node [base,above=4mm of d](dbox) {a};

\tikzset{% define styles for the different boxes, adjust parameters to wish
  base/.style={align=center, minimum height=5mm, minimum width=10mm, draw=black,dotted},
	circ/.style={align=center, minimum width=5mm, circle, draw=black,thick},
	% coord node style is used for drawing the arrows i.e. invisible nodes
  coord/.style={coordinate},
}
% add circle to f in a
\node [circ,above=2mm of a](acirc){};

% add large box to two boxes in b
\node [base, minimum width=28mm,minimum height=9mm, above=0mm of b](bboxLarge){};

% add larger box to c and d
\node [base, minimum width=24mm, minimum height =10.5mm, above=0mm of c](cboxLarge){};
\node [base, minimum width=24mm, minimum height =10.5mm, above=0mm of d](dboxLarge){};

% add coordinates to allow for drawing of the arrows
\node [coord,left=7mm of acirc](c_a1){};
\node [coord,right=7mm of acirc](c_a2){};
\node [coord,left=6mm of bbox1](c_b1){};
\node [coord,right=6mm of bbox2](c_b2){};
\node [coord,left=10mm of cbox](c_c1){};
\node [coord,right=10mm of cbox](c_c2){};
\node [coord,left=10mm of dbox](c_d1){};
\node [coord,right=10mm of dbox](c_d2){};

% finally add the arrows
\begin{scope}[->, thick]
\draw (c_a1) -- (acirc.west);
\draw (acirc.east) -- (c_a2);

\draw (c_b1) -- (bbox1);
\draw (bbox1) -- (bbox2);
\draw (bbox2) -- (c_b2);

\draw (c_c1) -- (cbox);
\draw (cbox) -- (c_c2);
\draw ([yshift=-4mm]c_c1) -- ([yshift=-4mm]c_c2);

\draw ([yshift=1mm]c_d1.east) -- ([yshift=1mm]dbox.west);
\draw ([yshift=1mm]dbox.east) -- ([yshift=1mm]c_d2.west);
\draw ([yshift=-1mm]dbox.east) to [out=346,in=194,looseness=6] ([yshift=-1mm]dbox.west);
\end{scope}

\end{tikzpicture}}
\caption{The core Arrow operators. Others, like second, may be built from these.}
\label{fig:arrowPic}
\end{figure}
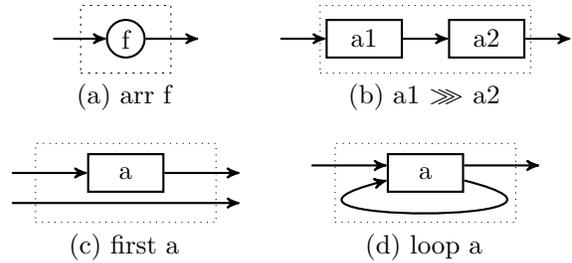

Arrowized FRP was introduced to plug a space leak in the original FRP work~\cite{hudakFRAN,liu2007plugging}.
By using the Arrow abstraction introduced in~\cite{hughes2000generalising}, which describes in a single type inputs and outputs, we can also describe reactive programs that process inputs and produce outputs over time.
At the top level, an Arrowized FRP program will have the form
\begin{equation*}
SF\ Input\ Output :: Signal\ Input \to Signal\ Output
\end{equation*}
which is a signal function type, parametrized by the type of input from the world and the type of output to the world.
The core Arrow operators, shown in Fig.~\ref{fig:arrowPic}, are used to composed multiple arrows into larger programs.

The abstractions used in different implementations of FRP vary in expressive power. Arrowized FRP has a smaller interface than a Monadic FRP~\cite{lindley2011idioms}, which prevents particular constructs that caused the aforementioned space leak.
This is also useful when choosing a core language to synthesize, as we will be able to simulate an Arrowized FRP program in most Monadic libraries.

\begin{figure}
\centering
\scalebox{1}{\begin{tikzpicture}[%
    >=stealth' % here you can choose the arrow tip of your choice
		]
\tikzset{% define styles for the different boxes, adjust parameters to wish
  base/.style={align=center, minimum height=5mm, minimum width=10mm, draw=black,thick},
	circ/.style={align=center, minimum width=10mm, circle, draw=black,thick},
	% coord node style is used for drawing the arrows i.e. invisible nodes
  coord/.style={coordinate},
}
\node [circ] (f) {f};
\node [coord,left=10mm of f.west] (c1) {};
\node [coord,right=20mm of f.east] (c2) {};
\node [coord, below=2.5mm of f.east] (c3) {};

\node [base,right=2.5mm of c3] (init){init i};

\node [coord, right=2.5mm of init.east] (p1){};
\node [coord, below=5mm of p1] (p2) {};
\node [coord, left=30mm of p2] (p3) {};
\node [coord, above=5mm of p3] (p4) {};

\begin{scope}[->, thick]
\draw ([yshift=2.5mm]c1) -- ([yshift=2.5mm,xshift=0.9mm]f.west);
\draw ([yshift=2.5mm,xshift=-0.9mm]f.east) -- ([yshift=2.5mm]c2);

\draw ([yshift=-2.5mm,xshift=-0.9mm]f.east) -- (init.west);

\draw (init.east) -- (p1) -- (p2) -- (p3) -- (p4) -- ([yshift=-2.5mm,xshift=0.9mm]f.west);
\end{scope}

\end{tikzpicture}}
\caption{loopD~~i~~f : a special loop from CCA that is always initialized with a user provided value i.}
\label{fig:arrowPic2}
\end{figure}
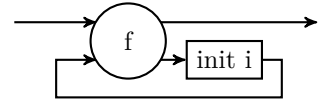

\paragraph{CCA}
We target a restricted set of arrows called Causal Commutative Arrows (CCA)~\cite{yallop2016causal,jfp/LiuCH11}.
Specifically, CCA adds additional laws to arrows that constrain their behavior and the type of state they may retain.
Of particular interest to our application is that CCA also introduces a special initialization operator, \texttt{init}.
This \texttt{init} operator allows for \texttt{loopD}, which is a loop that includes an initialization, as shown in Fig.~\ref{fig:arrowPic2}.

We use CCA as a minimal language for synthesis.
Our synthesis is able to support any FRP library which is at least as powerful as CCA.
Because CCA is again restricted in its interface, there are more libraries that can simulate CCA FRP than Arrowized FRP in general.
This makes our synthesis procedure possible for an even wider range of application scenarios.
We revisit the implications of CCA as our choice of computation abstraction in Sec.~\ref{sec:synthProp}.

\subsection{Reactive Synthesis}
\label{sec_rs}

The synthesis of a reactive system concerns the process of automatically
generating an implementation from a high level specification. The
reactive system acts as a deterministic controller, which reads inputs
and produces outputs over an infinite amount of time. In contrast,
a specification defines all input/output behavior pairs which are
valid, i.e., allowed to be produced by the controller.

In the classical synthesis setting, time is discrete and inputs and
outputs are given as vectors of Boolean signals. The standard
abstraction treats inputs and outputs as atomic propositions
$ \mathcal{I} \cup \mathcal{O} $, while their Boolean combinations
form an alphabet~$ \Sigma = 2^{\mathcal{I} \cup \mathcal{O}} $ of
alphabet symbols.

This fixes the behavior of the system to infinite sequences
$ \sigma = \sigma_{0}\sigma_{1}\sigma_{2} \ldots $ of alphabet
symbols~$ \sigma_{t} $. At every time~$ t $ signals appearing in the
set~$ \sigma_{t} $ are enabled (\name{true}), while signals not
in $ \sigma_{t} $ are disabled (\name{false}). The set of all
such sequences is denoted by~$ \Sigma^{\omega} $, where the
$ \omega $-operator induces the infinite concatenation of alphabet
symbols of $ \Sigma $. A deterministic solution links exactly one
output sequence $ \beta \in (2^{\mathcal{O}})^{\omega} $ to every
possible infinite sequence of inputs
$ \alpha \in (2^{\mathcal{I}})^{\omega} $, i.e., it is a total
function~$ f \colon (2^{\mathcal{I}})^{\omega} \to
(2^{\mathcal{O}})^{\omega} $. A specification describes an arbitrary
relation between input
sequences~\mbox{$ \alpha \in (2^{\mathcal{I}})^{\omega} $} and output
sequences~\mbox{$ \beta \in (2^{\mathcal{O}})^{\omega} $}.

\subsection{Connections between FRP and Reactive Systems}
\label{sec_connections}

A first inspection reveals that FRP fits into the definition of a reactive system, as given in Section~\ref{sec_rs}: an FRP program reads an infinite stream of input signals and finally produces a corresponding infinite output stream.
Nevertheless, FRP does not fit into the classical setting used for reactive systems, as the input and output streams in FRP are allowed to have arbitrary types.

To solve this problem, one could restrict FRP to just streams of enumerative types, which could then be reduced to a Boolean representation.
However, this would drop the necessity of almost all interesting features of FRP and it is questionable whether this restricted notion of FRP would give any benefits against \mbox{Mealy}/\mbox{Moore} automata or circuits, which are already used for reactive systems.
Additionally, it just creates an exponential blowup and does not provide any insights into the core problem.

Hence, it is more reasonable and interesting to ask whether it is possible to natively handle streams of arbitrary type within reactive systems.
Recall that FRP includes functional behavior, defined using standard functional paradigms, but also a control structure, defined via arrows and loops.
We will target synthesis of the control structure, leaving the functional level synthesis to tools such as \textsc{Myth}~\cite{osera2015type, kuncak2010complete} or \textsc{Synquid}~\cite{DBLP:conf/pldi/PolikarpovaKS16}.

\subsection{Temporal Stream Logic}

We use Temporal Stream Logic (TSL) for the specification of the
control flow behavior of functional reactive programs~\cite{tsl}. The
logic has been especially designed for synthesis and describes control
flow properties with respect to their temporal behavior over time. If
a TSL specification is realizable it can be turned into a Control Flow
Model (CFM), which is an abstract representation of the FRP network
covering all possible behavior switches.

Temporal Stream Logic builds on the notion of \mbox{\textit{updates}},
such as \mbox{$ \upd{\name{y}}{\name{f}~\name{x}} $}, which expresses
that the result of mapping the pure function~\name{f} to some input
stream~$ \name{x} $ is piped to the output stream~$ \name{y} $. The
execution of an update then is coupled with predicate evaluations
guiding the control flow decisions of the network. In combination with
Boolean and temporal operators, the logic allows for expressing even
complex temporally evolving FRP networks using only a short, but
precise description of the temporal behavior.

A useful advantage of TSL in contrast to other specification logics is
that function and predicate names, as used by the specification, are
only considered as symbolic literals. Therefore, the logic guarantees
that synthesized systems satisfy the specified behavior for all
possible implementations of these function and predicate literals. The
literals are still classified according to their arity, i.e., the
number of other function terms they are applied to, as well as by
their type: input, output, cell, function or predicate. Thus,
they can be considered similar to a function, passed as an argument, of
polymorphic type.

However, using TSL also comes with another important advantage. As
updates can be considered as pure building blocks lifted to the
temporal domain, the synthesis engine guarantees that a created CFM is
implementable using a static network. This especially is in contrast
to the well-known \texttt{switch} operation used by many FRP
libraries, which allows for the creation of dynamically evolving
networks. While \texttt{switch} is a very expressive operation in the
first place, it also comes with drawbacks. First, dynamically evolving
networks cannot provide run-time guarantees for memory requirements in
general, while static networks do. Second, the behavior of a
dynamically evolving network is hard to grasp in general, which
especially makes them unamenable for verification. Third, the use of
dynamic networks is out of scope for FRP applications with restricted
resources, as for example applications that are executed on embedded
devices \cite{DBLP:conf/aosd/SawadaW16} or are implemented directly in
hardware~\cite{clash2015}. Finally, while \texttt{switch} is an expressive
operation, it is not necessary for most applications~\cite{Winograd-Cort:2014}.
Thus, by using TSL as a specification language, we can avoid
these problems implicitly, as the logic guarantees the creation of a
static network.

\begin{figure}[t]
  \centering
  \begin{tikzpicture}[scale=0.8]

    \node[anchor=east,inner sep=0pt] at (-2.8,-1.5) {
      \small
      \begin{tabular}{c}
        inputs: \\[0.1em] $ \inames $
      \end{tabular}
    };

    \node at (0,1.48) {
      \small
      cells: $ \cells $
    };

    \node[anchor=west,inner sep=0pt] at (2.8,-1.5) {
      \small
      \begin{tabular}{c}
        outputs: \\[0.1em]
        $ \onames $
      \end{tabular}
    };

    \node at (0,0) {
      \begin{tikzpicture}[xscale=0.8,yscale=0.56]
        \node[fill, fill=blue!30,minimum height=5.5em, minimum width=10.5em] (C) {};

        \node at (C) {
          \small
          \begin{tabular}{c}
            \textit{reactive system} \\[0.4em]
            \textit{implementing a} \\[0.4em]
            \textit{TSL specification~$ \varphi $}
            \end{tabular}
        };

        \node[minimum size=0.9em] (H0) at (0,2.05) {};
        \node[minimum size=0.9em] (H1) at (0,2.7) {};
        \node[minimum size=0.9em] (H2) at (0,3.9) {};

        % signals
        \path[->,>=stealth,line width=0.7pt]
        ($ (C.west) + (-0.6,-1.3) $) edge ($ (C.west) + (0,-1.3) $)
        ($ (C.west) + (-0.6,-0.6) $) edge ($ (C.west) + (0,-0.6) $)
        ($ (C.west) + (-0.6,-0.3) $) edge ($ (C.west) + (0,-0.3) $)
        ($ (C.east) + (0,-1.3) $) edge ($ (C.east) + (0.6,-1.3) $)
        ($ (C.east) + (0,-0.6) $) edge ($ (C.east) + (0.6,-0.6) $)
        ($ (C.east) + (0,-0.3) $) edge ($ (C.east) + (0.6,-0.3) $)
        ;

        \node at ($ (C.west) + (-0.35,-0.85) $) {\scalebox{0.6}{$ \vdots $}};
        \node at ($ (C.east) + (0.25,-0.85) $) {\scalebox{0.6}{$ \vdots $}};

        % cells
        \draw[line width=0.7pt,-,>=stealth,gray]
        ($ (C.east) + (0,1.3) $) -- (2.7,1.3) |- (H0);
        \draw[line width=0.7pt,->,>=stealth,gray]
        (H0) -| (-2.7,1.3) -- ($ (C.west) + (0,1.3) $);

        \draw[line width=0.7pt,-,>=stealth,gray]
        ($ (C.east) + (0,1) $) -- (3,1) |- (H1);
        \draw[line width=0.7pt,->,>=stealth,gray]
        (H1) -| (-3,1) |- ($ (C.west) + (0,1) $);

        \draw[line width=0.7pt,-,>=stealth,gray]
        ($ (C.east) + (0,0.3) $) -- (3.6,0.3) |- (H2);
        \draw[line width=0.7pt,->,>=stealth,gray]
        (H2) -| (-3.6,0.3) |- ($ (C.west) + (0,0.3) $);

        \node at ($ (C.west) + (-0.35,0.75) $) {\scalebox{0.6}{$ \vdots $}};
        \node at ($ (C.east) + (0.25,0.75) $) {\scalebox{0.6}{$ \vdots $}};

        \fill[fill=orange!60]
        ($ (H0.north west) + (0,-0.1) $) --
        ($ (H0.south west) + (0,0.1) $) --
        ($ (H0.south west) + (0.1,0) $) --
        ($ (H0.south east) + (-0.1,0) $) --
        ($ (H0.south east) + (0,0.1) $) --
        ($ (H0.north east) + (0,-0.1) $) --
        ($ (H0.north east) + (-0.1,0) $) --
        ($ (H0.north west) + (0.1,0) $) --
        cycle;

        \fill[fill=orange!60]
        ($ (H1.north west) + (0,-0.1) $) --
        ($ (H1.south west) + (0,0.1) $) --
        ($ (H1.south west) + (0.1,0) $) --
        ($ (H1.south east) + (-0.1,0) $) --
        ($ (H1.south east) + (0,0.1) $) --
        ($ (H1.north east) + (0,-0.1) $) --
        ($ (H1.north east) + (-0.1,0) $) --
        ($ (H1.north west) + (0.1,0) $) --
        cycle;

        \fill[fill=orange!60]
        ($ (H2.north west) + (0,-0.1) $) --
        ($ (H2.south west) + (0,0.1) $) --
        ($ (H2.south west) + (0.1,0) $) --
        ($ (H2.south east) + (-0.1,0) $) --
        ($ (H2.south east) + (0,0.1) $) --
        ($ (H2.north east) + (0,-0.1) $) --
        ($ (H2.north east) + (-0.1,0) $) --
        ($ (H2.north west) + (0.1,0) $) --
        cycle;

      \end{tikzpicture}
    };
  \end{tikzpicture}
  \vspace{-1em}
  \caption{TSL System Architecture \mbox{\ }}
  \label{fig:architecture}
\end{figure}
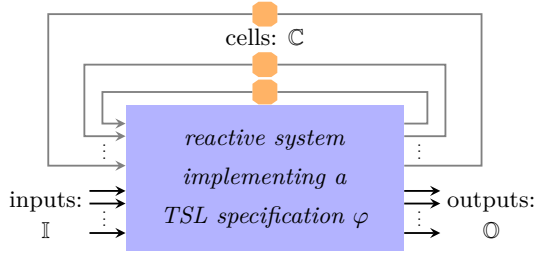

TSL specifications implicitly induce an architecture as shown in
\cref{fig:architecture}. As a basis, the syntax of TSL uses a term
based notion, build from input streams~$ \name{i} \in \inames $,
output streams~\mbox{$ \name{o} \in \onames $}, memory
cells~$ \name{c} \in \cells $, and function and predicate literals
$ \name{f} \in \fnames $ and $ \name{p} \in \pnames $ with
$ \pnames \subseteq \fnames $, respectively. The purpose of cells is
to memorize data values, output to a cell at time~$ t \in \dtime $, to
provide them again as inputs at time~$ t + 1 $. We differentiate
between function terms~$ \fterm \in \fterms $ and predicate
terms~$ \pterm \in \pterms $, build according to the following grammar
\begin{equation*}
  \begin{array}{rl}
  \fterm \ \;:= & \name{s}_{\name{i}} \sep \name{f}~\;\fterm^{0}~\;\fterm^{1}~\;\cdots~\;\fterm^{n-1} \\[0.3em]
  \pterm \ \;:= & \name{p}~\;\fterm^{0}~\;\fterm^{1}~\;\ldots~\;\fterm^{n-1}
  \end{array}
\end{equation*}
where $ \name{s}_{\name{i}} \in \inames \cup \cells $ is either an
input stream or a cell. In a TSL formula~$ \varphi $, predicate and
function terms are then combined to updates and with Boolean
connectives and temporal operators
\begin{equation*}
  \varphi \ := \ \pterm \ \ | \ \ \upd{\name{s}_{\name{o}}}{\fterm} \ \ | \ \ \varphi \wedge \varphi \ \ | \ \ \LTLnext \varphi \ \ | \ \ \varphi \LTLuntil \varphi
\end{equation*}
where $ \name{s}_{\name{o}} \in \onames \cup \cells $ is either an output signal or a cell.

To give semantics to TSL formulas~$ \varphi $ a universally quantified assignment
function~$ \assign{\cdot} \from \fnames \to \functions $ is used,
fixing an implementation for each predicate and function literal, as
well as input streams~$ \iota \from \inames \to \signals $. We will
only give an intuitive description of the semantics here. For a fully
formal definition the interested reader is referred to
\cite{tsl}. Intuitively, the semantics of TSL can be summarized as
follows:

\smallskip

\textbf{Predicate terms} are evaluated to either \texttt{true} or
\texttt{false}, by first selecting implementations for all function
and predicate literals according to $ \assign{\cdot} $, and then
applying them to inputs, as given by $ \iota $, and cells, using the
stored value at the current time~$ t $. The content of a cell thereby
is fixed iteratively by selecting the past values piped into the cell
over time. Cells are initialized using a special constant
provided as part of~$ \assign{\cdot} $.

\textbf{Function terms} are evaluated similar to predicate terms,
except that they can evaluate to any value of arbitrary type.

\textbf{Updates} are used to pipe the result of function term
evaluations to output streams or cells. Note that updates, as they
appear in a TSL formula, semantically are typed as Boolean
expressions. In that sense, update expressions are used in TSL formula
to state that a specific flow is used at a specific point in time,
where the expression evaluates to \texttt{true} if it is used and to
\texttt{false}, otherwise.

\begin{figure}
  \begin{tikzpicture}[scale=1.05]
    \node[anchor=north west] at (0,0) {

  \begin{tikzpicture}[anchor=center]
     \node[inner sep=0pt,minimum size=3mm,minimum width=7mm] (S) at (-1,0) {$ \cdots $};

     \node[draw, inner sep=0pt,minimum size=3mm,minimum width=7mm,
     fill=green!80!black] (A) at (0,0) {\tiny $ t $};

     \node[draw, inner sep=0pt,minimum
     size=3mm,minimum width=7mm,fill=green!80!black] (B) at (1,0) {\tiny $ t + 1 $};

     \node[inner sep=0pt,minimum
     size=3mm,minimum width=7mm] (C) at (2,0) {$ \cdots $};

     \path[->,>=stealth]
       (S) edge (A)
       (A) edge[blue!80,ultra thick] (B)
       (B) edge (C)
       ;

     \node[inner sep=1pt,rounded corners=1,fill=orange!50] (P0) at (0,0.5) {\tiny $ \LTLnext \varphi $};
     \node[inner sep=2pt] (P3) at (1,0.5) {\tiny $ \varphi $};

     \path (P0) edge[blue!80,->,ultra thick] (A)
     (P3) edge[green!50!black,->,ultra thick] (B);
   \end{tikzpicture}
 };

   \node[anchor=north west,yshift=-0.7] at (3.98,0) {

  \begin{tikzpicture}[anchor=center]
     \node[inner sep=0pt,minimum size=3mm,minimum width=7mm] (S) at (-1,0) {$ \cdots $};

     \node[draw, inner sep=0pt,minimum size=3mm,minimum width=7mm,
     fill=green!80!black] (A) at (0,0) {\tiny $ t $};

     \node[draw, inner sep=0pt,minimum
     size=3mm,minimum width=7mm,fill=green!80!black] (B) at (1,0) {\tiny $ t + 1 $};

     \node[inner sep=0pt,minimum
     size=3mm,minimum width=7mm] (C) at (2,0) {$ \cdots $};

     \path[->,>=stealth]
       (S) edge (A)
       (A) edge[blue!80,ultra thick] (B)
       (B) edge[blue!80,ultra thick] (C)
       ;

     \node[inner sep=1pt,rounded corners=1,fill=orange!50] (P0) at (0,0.5) {\tikz \node[inner sep=0pt,rounded corners=0]{\tiny $ \LTLglobally \varphi $};};
     \node[inner sep=2pt] (P0m) at (0,-0.5) {\tiny $ \varphi $};
     \node[inner sep=2pt] (P1m) at (1,-0.5) {\tiny $ \varphi $};

     \path (P0) edge[blue!80,->,ultra thick] (A)
     (P0m) edge[green!50!black,->,ultra thick] (A)
     (P1m) edge[green!50!black,->,ultra thick] (B);
   \end{tikzpicture}
   };
   \node[anchor=north west] at (0,-1.4) {

  \begin{tikzpicture}[anchor=center]
     \node[inner sep=0pt,minimum size=3mm,minimum width=7mm] (S) at (-1,0) {$ \cdots $};

     \node[draw, inner sep=0pt,minimum size=3mm,minimum width=7mm,
     fill=green!80!black] (A) at (0,0) {\tiny $ t $};

     \node[draw, inner sep=0pt,minimum
     size=3mm,minimum width=7mm,fill=green!80!black] (B) at (1,0) {\tiny $ t + 1 $};

     \node[inner sep=0pt,minimum
     size=3mm,minimum width=7mm] (C) at (2,0) {$ \cdots $};

     \node[draw, inner sep=0pt,minimum
     size=3mm,minimum width=7mm,fill=green!80!black] (D) at (3,0) {\tiny $ t' \!\! - \! 1 $};

     \node[draw, inner sep=0pt,minimum
     size=3mm,minimum width=7mm,fill=green!80!black] (E) at (4,0) {\tiny $ t' $};

     \node[draw, inner sep=0pt,minimum
     size=3mm,minimum width=7mm,fill=blue!20] (F) at (5,0) {\tiny $ t'\! + 1 $};

     \node[inner sep=0pt,minimum size=3mm,minimum width=7mm] (G) at (6,0) {$ \cdots $};

     \path[->,>=stealth]
       (S) edge (A)
       (A) edge[blue!80,ultra thick] (B)
       (B) edge[blue!80,ultra thick] (C)
       (C) edge[blue!80,ultra thick] (D)
       (D) edge[blue!80,ultra thick] (E)
       (E) edge (F)
       (F) edge (G)
       ;

     \node[inner sep=1pt,rounded corners=1,fill=orange!50] (P0) at (0,0.5) {\tikz \node[inner sep=0pt,rounded corners=0]{\tiny $ \LTLfinally \varphi  $};};
     \node[inner sep=2pt] (P3) at (4,0.5) {\tiny $ \varphi $};

     \path (P0) edge[blue!80,->,ultra thick] (A)
     (P3) edge[green!50!black,->,ultra thick] (E);
   \end{tikzpicture}
   };
   \node[anchor=north west] at (0,-2.4) {

  \begin{tikzpicture}[anchor=center]
     \node[inner sep=0pt,minimum size=3mm,minimum width=7mm] (S) at (-1,0) {$ \cdots $};

     \node[draw, inner sep=0pt,minimum size=3mm,minimum width=7mm,
     fill=green!80!black] (A) at (0,0) {\tiny $ t $};

     \node[draw, inner sep=0pt,minimum
     size=3mm,minimum width=7mm,fill=green!80!black] (B) at (1,0) {\tiny $ t + 1 $};

     \node[inner sep=0pt,minimum
     size=3mm,minimum width=7mm] (C) at (2,0) {$ \cdots $};

     \node[draw, inner sep=0pt,minimum
     size=3mm,minimum width=7mm,fill=green!80!black] (D) at (3,0) {\tiny $ t' \!\! - \! 1 $};

     \node[draw, inner sep=0pt,minimum
     size=3mm,minimum width=7mm,fill=green!80!black] (E) at (4,0) {\tiny $ t' $};

     \node[draw, inner sep=0pt,minimum
     size=3mm,minimum width=7mm,fill=blue!20] (F) at (5,0) {\tiny $ t'\! + 1 $};

     \node[inner sep=0pt,minimum size=3mm,minimum width=7mm] (G) at (6,0) {$ \cdots $};

     \path[->,>=stealth]
       (S) edge (A)
       (A) edge[blue!80,ultra thick] (B)
       (B) edge[blue!80,ultra thick] (C)
       (C) edge[blue!80,ultra thick] (D)
       (D) edge[blue!80,ultra thick] (E)
       (E) edge (F)
       (F) edge (G)
       ;

     \node[inner sep=1pt,rounded corners=1,fill=orange!50] (P0) at (0,0.5) {\tiny $ \psi \LTLuntil \varphi $};
     \node[inner sep=2pt] (P3) at (4,0.5) {\tiny $ \varphi $};
     \node[inner sep=2pt] (P0m) at (0,-0.5) {\tiny $ \psi $};
     \node[inner sep=2pt] (P1m) at (1,-0.5) {\tiny $ \psi $};
     \node[inner sep=2pt] (P2m) at (3,-0.5) {\tiny $ \psi $};

     \path (P0) edge[blue!80,->,ultra thick] (A)
     (P3) edge[green!50!black,->,ultra thick] (E)
     (P0m) edge[green!50!black,->,ultra thick] (A)
     (P1m) edge[green!50!black,->,ultra thick] (B)
     (P2m) edge[green!50!black,->,ultra thick] (D);
   \end{tikzpicture}
   };
   \node[anchor=north west] at (0,-3.8) {

  \begin{tikzpicture}[anchor=center]
     \node[inner sep=0pt,minimum size=3mm,minimum width=7mm] (S) at (-1,0) {$ \cdots $};

     \node[inner sep=0pt,minimum size=3mm,minimum width=7mm] (A) at (0,0) {};

     \fill[green!80!black] (A.west) -- (A.north west) -- (A.north east) -- (A.east) -- cycle;
     \fill[yellow!80!black] (A.west) -- (A.south west) -- (A.south east) -- (A.east) --  cycle;

     \node[draw, inner sep=0pt,minimum size=3mm,minimum width=7mm] (A) at (0,0) {\tiny $ t $};

     \node[inner sep=0pt,minimum size=3mm,minimum width=7mm] (B) at (1,0) {};

     \fill[green!80!black] (B.west) -- (B.north west) -- (B.north east) -- (B.east) -- cycle;
     \fill[yellow!80!black] (B.west) -- (B.south west) -- (B.south east) -- (B.east) --  cycle;

     \node[draw, inner sep=0pt,minimum
     size=3mm,minimum width=7mm] (B) at (1,0) {\tiny $ t + 1 $};

     \node[inner sep=0pt,minimum
     size=3mm,minimum width=7mm] (C) at (2,0) {$ \cdots $};

     \node[inner sep=0pt,minimum size=3mm,minimum width=7mm] (D) at (3,0) {};

     \fill[green!80!black] (D.west) -- (D.north west) -- (D.north east) -- (D.east) -- cycle;
     \fill[yellow!80!black] (D.west) -- (D.south west) -- (D.south east) -- (D.east) --  cycle;

     \node[draw, inner sep=0pt,minimum
     size=3mm,minimum width=7mm] (D) at (3,0) {\tiny $ t' \!\! - \! 1 $};

     \node[inner sep=0pt,minimum size=3mm,minimum width=7mm] (E) at (4,0) {};

     \fill[green!80!black] (E.west) -- (E.north west) -- (E.north east) -- (E.east) -- cycle;
     \fill[yellow!80!black] (E.west) -- (E.south west) -- (E.south east) -- (E.east) --  cycle;

     \node[draw, inner sep=0pt,minimum
     size=3mm,minimum width=7mm] (E) at (4,0) {\tiny $ t' $};

     \node[draw, inner sep=0pt,minimum
     size=3mm,minimum width=7mm,fill=yellow!80!black] (F) at (5,0) {\tiny $ t'\! + 1 $};

     \node[inner sep=0pt,minimum size=3mm,minimum width=7mm] (G) at (6,0) {$ \cdots $};

     \path[->,>=stealth]
       (S) edge (A)
       (A) edge[blue!80,ultra thick] (B)
       (B) edge[blue!80,ultra thick] (C)
       (C) edge[blue!80,ultra thick] (D)
       (D) edge[blue!80,ultra thick] (E)
       (E) edge[blue!80,ultra thick] (F)
       (F) edge[blue!80,ultra thick] (G)
       ;

     \node[inner sep=1pt,rounded corners=1,fill=orange!50] (P0) at (0,0.5) {\tiny $ \psi \LTLrelease \varphi $};
     \node[inner sep=2pt] (P3) at (4,0.5) {\tiny $ \psi $};
     \node[inner sep=2pt] (P0m) at (0,-0.5) {\tiny $ \varphi $};
     \node[inner sep=2pt] (P1m) at (1,-0.5) {\tiny $ \varphi $};
     \node[inner sep=2pt] (P2m) at (3,-0.5) {\tiny $ \varphi $};
     \node[inner sep=2pt] (P3m) at (4,-0.5) {\tiny $ \varphi $};
     \node[inner sep=2pt] (P4m) at (5,-0.5) {\tiny $ \varphi $};

     \path (P0) edge[blue!80,->,ultra thick] (A)
     (P3) edge[green!50!black,->,ultra thick] (E)
     (P0m) edge[yellow!50!black,->,ultra thick] (A)
     (P1m) edge[yellow!50!black,->,ultra thick] (B)
     (P2m) edge[yellow!50!black,->,ultra thick] (D)
     (P3m) edge[yellow!50!black,->,ultra thick] (E)
     (P4m) edge[yellow!50!black,->,ultra thick] (F);

     \begin{scope}
       \clip (P0m.south) -- (A.north) -- ++(0.7,0) |- cycle;
       \path (P0m) edge[green!50!black,->,ultra thick] (A);
     \end{scope}
     \begin{scope}
       \clip (P1m.south) -- (B.north) -- ++(0.7,0) |- cycle;
       \path (P1m) edge[green!50!black,->,ultra thick] (B);
     \end{scope}
     \begin{scope}
       \clip (P2m.south) -- (D.north) -- ++(0.7,0) |- cycle;
       \path (P2m) edge[green!50!black,->,ultra thick] (D);
     \end{scope}
     \begin{scope}
       \clip (P3m.south) -- (E.north) -- ++(0.7,0) |- cycle;
       \path (P3m) edge[green!50!black,->,ultra thick] (E);
     \end{scope}
   \end{tikzpicture}
   };
   \node[anchor=north west] at (0,-5.2) {
  \begin{tikzpicture}[anchor=center]
     \node[inner sep=0pt,minimum size=3mm,minimum width=7mm] (S) at (-1,0) {$ \cdots $};

     \node[inner sep=0pt,minimum size=3mm,minimum width=7mm] (A) at (0,0) {};

     \fill[green!80!black] (A.west) -- (A.north west) -- (A.north east) -- (A.east) -- cycle;
     \fill[yellow!80!black] (A.west) -- (A.south west) -- (A.south east) -- (A.east) --  cycle;

     \node[draw, inner sep=0pt,minimum size=3mm,minimum width=7mm] (A) at (0,0) {\tiny $ t $};

     \node[inner sep=0pt,minimum size=3mm,minimum width=7mm] (B) at (1,0) {};

     \fill[green!80!black] (B.west) -- (B.north west) -- (B.north east) -- (B.east) -- cycle;
     \fill[yellow!80!black] (B.west) -- (B.south west) -- (B.south east) -- (B.east) --  cycle;

     \node[draw, inner sep=0pt,minimum
     size=3mm,minimum width=7mm] (B) at (1,0) {\tiny $ t + 1 $};

     \node[inner sep=0pt,minimum
     size=3mm,minimum width=7mm] (C) at (2,0) {$ \cdots $};

     \node[inner sep=0pt,minimum size=3mm,minimum width=7mm] (D) at (3,0) {};

     \fill[green!80!black] (D.west) -- (D.north west) -- (D.north east) -- (D.east) -- cycle;
     \fill[yellow!80!black] (D.west) -- (D.south west) -- (D.south east) -- (D.east) --  cycle;

     \node[draw, inner sep=0pt,minimum
     size=3mm,minimum width=7mm] (D) at (3,0) {\tiny $ t' \!\! - \! 1 $};

     \node[inner sep=0pt,minimum size=3mm,minimum width=7mm] (E) at (4,0) {};

     \fill[green!80!black] (E.west) -- (E.north west) -- (E.north east) -- (E.east) -- cycle;
     \fill[yellow!80!black] (E.west) -- (E.south west) -- (E.south east) -- (E.east) --  cycle;

     \node[draw, inner sep=0pt,minimum
     size=3mm,minimum width=7mm] (E) at (4,0) {\tiny $ t' $};

     \node[draw, inner sep=0pt,minimum
     size=3mm,minimum width=7mm,fill=yellow!80!black] (F) at (5,0) {\tiny $ t'\! + 1 $};

     \node[inner sep=0pt,minimum size=3mm,minimum width=7mm] (G) at (6,0) {$ \cdots $};

     \path[->,>=stealth]
       (S) edge (A)
       (A) edge[blue!80,ultra thick] (B)
       (B) edge[blue!80,ultra thick] (C)
       (C) edge[blue!80,ultra thick] (D)
       (D) edge[blue!80,ultra thick] (E)
       (E) edge[blue!80,ultra thick] (F)
       (F) edge[blue!80,ultra thick] (G)
       ;

     \node[inner sep=1pt,rounded corners=1,fill=orange!50] (P0) at (0,0.5) {\tiny $ \psi \LTLweakuntil \varphi $};
     \node[inner sep=2pt] (P3) at (4,0.5) {\tiny $ \varphi $};
     \node[inner sep=2pt] (P0m) at (0,-0.5) {\tiny $ \psi $};
     \node[inner sep=2pt] (P1m) at (1,-0.5) {\tiny $ \psi $};
     \node[inner sep=2pt] (P2m) at (3,-0.5) {\tiny $ \psi $};
     \node[inner sep=2pt] (P3m) at (4,-0.5) {\tiny $ \psi $};
     \node[inner sep=2pt] (P4m) at (5,-0.5) {\tiny $ \psi $};

     \path (P0) edge[blue!80,->,ultra thick] (A)
     (P3) edge[green!50!black,->,ultra thick] (E)
     (P0m) edge[yellow!50!black,->,ultra thick] (A)
     (P1m) edge[yellow!50!black,->,ultra thick] (B)
     (P2m) edge[yellow!50!black,->,ultra thick] (D)
     (P3m) edge[yellow!50!black,->,ultra thick] (E)
     (P4m) edge[yellow!50!black,->,ultra thick] (F);

     \begin{scope}
       \clip (P0m.south) -- (A.north) -- ++(0.7,0) |- cycle;
       \path (P0m) edge[green!50!black,->,ultra thick] (A);
     \end{scope}
     \begin{scope}
       \clip (P1m.south) -- (B.north) -- ++(0.7,0) |- cycle;
       \path (P1m) edge[green!50!black,->,ultra thick] (B);
     \end{scope}
     \begin{scope}
       \clip (P2m.south) -- (D.north) -- ++(0.7,0) |- cycle;
       \path (P2m) edge[green!50!black,->,ultra thick] (D);
     \end{scope}
   \end{tikzpicture}
 };
 \draw (0,-1.44) -- ++(7.8,0);
 \draw (0,-2.44) -- ++(7.8,0);
 \draw (0,-3.84) -- ++(7.8,0);
 \draw (0,-5.24) -- ++(7.8,0);

 \draw (3.94,-1.24) -- ++(0,1);

\end{tikzpicture}
\vspace{-1.2em}
\caption{Temporal behavior of the operators \textit{next},
  \textit{always}, \textit{finally}, \textit{until}, \textit{release},
  and \textit{weak until}. In case of \textit{release} and
  \textit{weak until}, the formula is either fulfilled by satisfying
  the top behavior (green) or the bottom behavior (yellow). The blue
  arrows on the time axis indicate the temporal scope of
  the operators over time.}
\label{fig:temporal}
\end{figure}
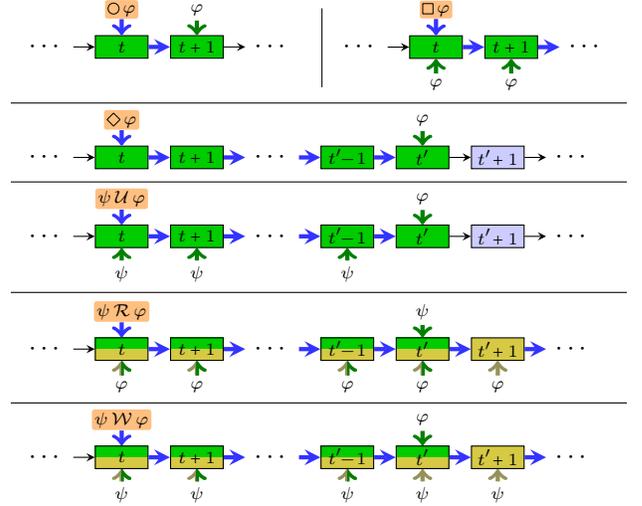

The \textbf{Boolean operators} \textit{negation}~[$ \neg $] and
\textit{con\-junc-\linebreak tion}~[$ \wedge $], and the \textbf{temporal operators}
\textit{next}~[$ \mathop{\LTLnext} $] and
\textit{until}~[$ \hspace{0.6pt}\mathop{\LTLuntil} \hspace{0.2pt}$]
have standard semantics.
Intuitive behavior for the temporal operators is given in
\cref{fig:temporal}, including also the derived operators
\textit{release}~[$ \varphi \LTLrelease \psi \equiv \neg
$(($\neg \psi$)$ \LTLuntil $($\neg \varphi$))$ $],
\textit{finally}~[$ \LTLfinally \varphi \equiv \emph{true} \LTLuntil
\varphi $],
\textit{always}~[$ \LTLglobally \varphi \equiv \emph{false}
\LTLrelease \varphi $], and the \textit{weak} version of
\textit{until}
[$ \varphi \LTLweakuntil \psi \equiv (\varphi \LTLuntil \psi) \vee
(\LTLglobally \varphi) $].

The synthesis problem of creating a CFM~$ \cfm $  satisfying a TSL
specification~$ \varphi $ then is formalized by
\begin{equation*}
  \exists \cfm. \ \, \forall \iota. \ \, \forall \assign{\cdot}. \ \, \branch{\cfm}{\iota}, \iota \sats \varphi
\end{equation*}
where $ \branch{\cfm}{\iota} $ denotes the output produced by $ \cfm $
under the input $ \iota $. Note that the CFM~$ \cfm $ must satisfy the
specification for all possibly chosen function and predicate
implementations, as selected by~$ \assign{\cdot} $, and all possible
inputs~$ \iota $, which is also the reason for the synthesis problem being
undecidable in general.

\begin{theorem}[\cite{tsl}]
  The synthesis problem of TSL is undecidable.
\end{theorem}

\section{System Design with TSL}
\label{sec:motiv}

\begin{figure}
  \centering

  \begin{tikzpicture}
    \node at (5,0) {
      \begin{tikzpicture}
        \draw[gray,rounded corners=10,fill=gray!70] (-1.8,-2.4) rectangle (1.8,2.4);

        \draw[gray, fill=green!20!yellow] (-1.3,0) rectangle (1.3,1.9);

        \node[draw,gray,fill=white,circle,minimum size=2.3em] (S) at (0,-0.8) {\color{black}\scriptsize \texttt{SEC}};
        \node[draw,gray,fill=white,circle,minimum size=2.3em] (M) at (-1,-0.8) {\color{black}\scriptsize \texttt{MIN}};
        \node[draw,gray,fill=white,circle,minimum size=2.3em,inner sep=-3pt] (A) at (1,-0.8) {\color{black}\tiny \begin{tabular}{c} \texttt{STOP} \\[0.1em] \texttt{START} \end{tabular}};
        \draw ($ (A.west) + (0.15,0) $) -- ($ (A.east) + (-0.15,0) $);
        \draw[line width=0.08em] (M.south) -- ++(0,-0.1) -- node[fill=gray!70,inner sep=1pt] {\tiny \texttt{RESET}} ($ (S.south) + (0,-0.1) $) -- (S.south);

        \node at (-0.9,0.8) {
          \begin{tikzpicture}
            \fill (0,0) -- (-0.05,0.05) -- (-0.05,0.55) -- (-0.02,0.58) -- (0.05,0.5) -- (0.05,0.05) -- cycle;
            \fill (0,0) -- (-0.05,-0.05) -- (-0.05,-0.55) -- (-0.02,-0.58) -- (0.05,-0.5) -- (0.05,-0.05) -- cycle;
            \fill (-0.005,0.585) -- (0.06,0.51) -- (0.35,0.51) -- (0.415,0.585) -- (0.39,0.61) -- (0.02,0.61) -- cycle;
            \fill (-0.005,-0.585) -- (0.06,-0.51) -- (0.35,-0.51) -- (0.415,-0.585) -- (0.39,-0.61) -- (0.02,-0.61) -- cycle;
            \fill (0.408,0) -- (0.458,0.05) -- (0.458,0.55) -- (0.428,0.58) -- (0.358,0.5) -- (0.358,0.05) -- cycle;
            \fill (0.408,0) -- (0.458,-0.05) -- (0.458,-0.55) -- (0.428,-0.58) -- (0.358,-0.5) -- (0.358,-0.05) -- cycle;
          \end{tikzpicture}
        };
        \node at (-0.35,0.8) {
          \begin{tikzpicture}
            \fill (0,0) -- (-0.05,0.05) -- (-0.05,0.55) -- (-0.02,0.58) -- (0.05,0.5) -- (0.05,0.05) -- cycle;
            \fill (0,0) -- (-0.05,-0.05) -- (-0.05,-0.55) -- (-0.02,-0.58) -- (0.05,-0.5) -- (0.05,-0.05) -- cycle;
            \fill (-0.005,0.585) -- (0.06,0.51) -- (0.35,0.51) -- (0.415,0.585) -- (0.39,0.61) -- (0.02,0.61) -- cycle;
            \fill (-0.005,-0.585) -- (0.06,-0.51) -- (0.35,-0.51) -- (0.415,-0.585) -- (0.39,-0.61) -- (0.02,-0.61) -- cycle;
            \fill (0.408,0) -- (0.458,0.05) -- (0.458,0.55) -- (0.428,0.58) -- (0.358,0.5) -- (0.358,0.05) -- cycle;
            \fill (0.408,0) -- (0.458,-0.05) -- (0.458,-0.55) -- (0.428,-0.58) -- (0.358,-0.5) -- (0.358,-0.05) -- cycle;
          \end{tikzpicture}
        };

        \node at (0.35,0.8) {
          \begin{tikzpicture}
            \fill (0,0) -- (-0.05,0.05) -- (-0.05,0.55) -- (-0.02,0.58) -- (0.05,0.5) -- (0.05,0.05) -- cycle;
            \fill (0,0) -- (-0.05,-0.05) -- (-0.05,-0.55) -- (-0.02,-0.58) -- (0.05,-0.5) -- (0.05,-0.05) -- cycle;
            \fill (-0.005,0.585) -- (0.06,0.51) -- (0.35,0.51) -- (0.415,0.585) -- (0.39,0.61) -- (0.02,0.61) -- cycle;
            \fill (-0.005,-0.585) -- (0.06,-0.51) -- (0.35,-0.51) -- (0.415,-0.585) -- (0.39,-0.61) -- (0.02,-0.61) -- cycle;
            \fill (0.408,0) -- (0.458,0.05) -- (0.458,0.55) -- (0.428,0.58) -- (0.358,0.5) -- (0.358,0.05) -- cycle;
            \fill (0.408,0) -- (0.458,-0.05) -- (0.458,-0.55) -- (0.428,-0.58) -- (0.358,-0.5) -- (0.358,-0.05) -- cycle;
          \end{tikzpicture}
        };

        \node at (0.9,0.8) {
          \begin{tikzpicture}
            \fill (0,0) -- (-0.05,0.05) -- (-0.05,0.55) -- (-0.02,0.58) -- (0.05,0.5) -- (0.05,0.05) -- cycle;
            \fill (0,0) -- (-0.05,-0.05) -- (-0.05,-0.55) -- (-0.02,-0.58) -- (0.05,-0.5) -- (0.05,-0.05) -- cycle;
            \fill (-0.005,0.585) -- (0.06,0.51) -- (0.35,0.51) -- (0.415,0.585) -- (0.39,0.61) -- (0.02,0.61) -- cycle;
            \fill (-0.005,-0.585) -- (0.06,-0.51) -- (0.35,-0.51) -- (0.415,-0.585) -- (0.39,-0.61) -- (0.02,-0.61) -- cycle;
            \fill (0.408,0) -- (0.458,0.05) -- (0.458,0.55) -- (0.428,0.58) -- (0.358,0.5) -- (0.358,0.05) -- cycle;
            \fill (0.408,0) -- (0.458,-0.05) -- (0.458,-0.55) -- (0.428,-0.58) -- (0.358,-0.5) -- (0.358,-0.05) -- cycle;
          \end{tikzpicture}
        };

        \node at (0.625,1.65) { \texttt{SEC} };
        \node at (-0.625,1.65) { \texttt{MIN} };
      \end{tikzpicture}
    };

    \node[anchor=west] (D) at (7,2.2) {\small \texttt{display}};

    \node[anchor=west] (S) at (7,1.48) {\small \texttt{seconds}};
    \node[anchor=west] (M) at (7,-0.12) {\small \texttt{minutes}};

    \node[anchor=west] (SA) at (7,-0.8) {\small \texttt{start/stop timer}};

    \node[anchor=west] (SS) at (7,-1.43) {\small \texttt{increase seconds}};

    \node[anchor=west] (SM) at (7,-2.08) {\small \texttt{increase minutes}};

    \coordinate (X) at (6.4,-0.8);
    \coordinate (Y) at (5.6,0);
    \coordinate (Z) at (3.7,-1.25);

    \draw[gray!50!black] (5,1.7) -- (5.5,2.2) -- (D);
    \draw[gray!50!black] (5.625,1.4) -- (5.7,1.48) -- (S);
    \draw[gray!50!black] (4.375,0.18) -- (4.7,-0.1) -- (M);
    \draw[gray!50!black] (X) -- (X -| SA.west);
    \draw[gray!50!black] (5.2,-1.15) -- (Y |- SS.west) -- (SS);
    \draw[gray!50!black] (3.8,-1.15) -- (Z) -- ($ (Z |- SM.west) + (0,0.2) $) -- ++(0.2,-0.2) -- (SM.west);
  \end{tikzpicture}
  \vspace{-0.5em}
  \caption{Timer application}
  \label{fig:timer}
\end{figure}
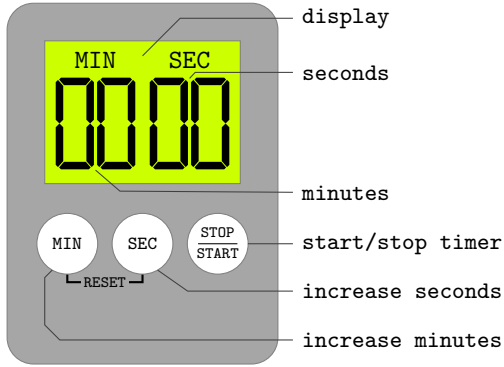

We demonstrate the advantages of using TSL as a specification
language for the development of FRP applications using the example
application of a kitchen timer, as presented in \cref{fig:timer}. The
timer consists of three buttons, a screen displaying the currently set
time and a buzzer to produce an alarm. The button values are provided
as Boolean input streams to the system, which deliver \texttt{true},
as long as a button is pressed, and \texttt{false} otherwise. In
addition, there is another input stream providing the time passed
since the last execution of the network, which is utilized to
synchronize the time displayed with the clock
of the application framework.

Similar to the button inputs, the system must output a Boolean data
stream to control the buzzer, which is turned on whenever the output
is \texttt{true}. The second output, the system must provide, delivers
the data to be displayed on the screen, where the data type is fixed
by the utilized application framework as well.

For our development plan, we consider the following list of
requirements to be implemented by the timer:
\begin{enumerate}

\item Whenever the \texttt{MIN} and \texttt{SEC} buttons are pressed
  simultaneously the timer is reset, meaning the time is set to zero
  and the system stays idle until the next button gets pressed.

\item If only the \texttt{MIN} button is pressed and the timer is not
  currently counting up or down, then the currently set time is
  increased by one minute.

\item If only the \texttt{SEC} button is pressed and the timer is not
  currently counting up or down, then the currently set time is
  increased by one second.

\item As long as no time greater than zero has been set and the system
  is idle: if the \texttt{START/STOP} button is pressed and the timer
  is not already counting up or down, then the timer starts counting
  up until it is stopped by any button pressed.

\item If a time has been set and the \texttt{START/STOP} button is
  pressed, while the timer is not currently counting up or down, then
  the timer starts counting down until it is stopped by any button
  pressed.

\item The timer can always be stopped by pressing any button while
  counting up or down.

\item It is possible to start the timer and to set some time
  simultaneously.

\item The buzzer beeps on any button press and after the counter
  reaches zero while counting down.  

\item The display always shows the time currently set.

\end{enumerate}
While it requires a certain amount of engineering to find the right
control behavior, especially for fixing the additional state to manage the
different modes, when directly implementing the application on top of
an FRP library, it is an easy task to specify the control behavior
with TSL. We first fix possible operations on \name{time}, which is used as
a cell for holding the currently set time.
\begin{equation*}
  \begin{array}{rcl}
    \textsc{countup}   & := & \upd{\name{time}}{\name{countup}~\name{time}~\name{dt}} \\[0.2em]
    \textsc{countdown} & := & \upd{\name{time}}{\name{countdown}~\name{time}~\name{dt}} \\[0.2em]
    \textsc{incmin}    & := & \upd{\name{time}}{\name{incMinutes}~\name{time}} \\[0.2em]
    \textsc{incsec}    & := & \upd{\name{time}}{\name{incSeconds}~\name{time}} \\[0.2em]
    \textsc{idle}      & := & \upd{\name{time}}{\name{time}} 
    \end{array}
\end{equation*}
The used literals~\name{countup}, \name{countdown}, \name{incMinutes},
and \name{incSeconds} represent pure functions that update the value
of \name{time} accordingly, while the input signal~\name{dt} delivers
the time difference since the last execution of the network. By the
semantics of TSL it is already ensured that assignments to the same
cell are always mutually exclusive, i.e., it can never be the case that time
is counting up and the minutes are increased in the same time step.

Next, we fix the properties of the behavior of \name{time} that
influence the control flow behavior. In our case, we need a predicate
to check whether the time currently set is zero or not
\begin{equation*}
  \begin{array}{rcl}  
    \textsc{zero} & := & \name{eq}~\name{time}~\name{zero}
  \end{array}
\end{equation*}  
where \name{zero} is a constant function of the same type as
\name{time}.
We also fix some sub-properties, that are useful to express conditions
regularly appearing in the main specification later. In our case these
are
\begin{equation*}
  \!\!\!\begin{array}{rcl}    
    \textsc{reset} & := & \name{btnMin} \, \wedge \, \name{btnSec} \\[0.2em]
    \textsc{counting} & := & \textsc{countup} \, \vee \, \textsc{countdown} \\[0.3em]
    \textsc{anykey} & := & \textit{press}~\name{btnMin} \, \vee \, \textit{press}~\name{btnSec} \\
    && \quad \vee \; \textit{press}~\name{btnStartStop} \\[0.2em]
    \textsc{start} & := & \textit{press}~\name{btnStartStop} \\
    && \quad \wedge \, \neg\textit{press}~\name{btnMin}
       \, \wedge \, \neg \textit{press}~\name{btnSec} \\[0.2em]
    \textsc{start\&min} & := & \textit{press}~\name{btnStartStop} \, \wedge \, \textit{press}~\name{btnMin} \\
    && \quad \wedge \, \LTLnext \text{(} \neg \name{btnSec} \, \wedge \, \LTLnext \neg \name{btnSec} \text{)}\\[0.2em]
    \textsc{start\&sec} & := & \textit{press}~\name{btnStartStop} \, \wedge \, \textit{press}~\name{btnSec} \\
    && \quad \wedge \, \LTLnext \text{(} \neg \name{btnMin} \, \wedge \, \LTLnext \neg \name{btnMin} \text{)}
  \end{array}
\end{equation*}
The literals \name{btnMin}, \name{btnSec}, and \name{btnStartStop} represent the
input signals for the three buttons, respectively. The function
\textit{press} is used as a helper function for improved readability
and is defined as
\begin{equation*}
  \begin{array}{rcl}      
    \textit{press}~x & := & \neg \, x \; \wedge \; \LTLnext x
  \end{array}
\end{equation*}
This is all we need to implement the invariants of the aforementioned
requirements:
\begin{equation*}
  \begin{array}{rcl}        
    \psi_{1} & := & \textsc{reset} \; \leftrightarrow \; \upd{\name{time}}{\name{zero}} \\[0.2em]
    \psi_{2} & := & \neg\textsc{counting} \,\wedge\, \textit{press}~\name{btnMin} \, \wedge \, \LTLnext \neg \name{btnSec} \\ 
                &&  \quad \leftrightarrow \; \LTLnext \, \textsc{incmin} \\[0.2em]
    \psi_{3} & := &  \neg\textsc{counting} \, \wedge \, \textit{press}~\name{btnSec} \,\wedge\, \LTLnext \neg \name{btnMin} \\
                && \quad \leftrightarrow \; \LTLnext \, \textsc{incsec} \\[0.2em]

    \psi_{4} & := & \textsc{zero} \rightarrow \\
             && \quad \big( \text{(} \, \textsc{idle} \, \wedge \, \textsc{start} \\
    && \qquad \quad \rightarrow \LTLnext \, \textit{tillAnyInput}~\textsc{countup} \text{)} \\[0.1em]
             && \qquad \qquad \ \ \LTLweakuntil \text{(} \textsc{incmin} \vee \textsc{incsec} \text{)} \big) \\[0.2em]

    \psi_{5} & := & \textsc{incmin} \,\vee \, \textsc{incsec} \rightarrow \\
             && \quad \big(\text{(}\neg \textsc{counting} \, \wedge \, \textsc{start} \\
             && \qquad \quad \rightarrow \LTLnext \, \textit{tillAnyInput}~\textsc{countdown} \text{)} \\
             && \qquad \qquad \ \ \LTLweakuntil \, \LTLnext \, \textsc{zero} \big) \\[0.2em]
    \psi_{6} & := & \textsc{counting} \, \wedge \, \textsc{anykey} \, \wedge \, \LTLnext \neg \textsc{reset} \\
             && \quad   \rightarrow \LTLnext \, \textit{tillAnyInput}~\textsc{idle} \\[0.2em]

    \psi_{7} & := & \neg \textsc{counting} \wedge (\textsc{start\&min} \, \vee \, \textsc{start\&sec}) \\
    && \quad \rightarrow \LTLnext \, \LTLnext \, \textit{tillAnyInput}~\textsc{countdown} \\[0.2em]

    \psi_{8} & := & \text{(} \upd{\name{beep}}{\name{true}} \oplus \upd{\name{beep}}{\name{false}} \text{)} \, \wedge \mbox{\ }\\[0.1em]
                 && \big( \!\LTLnext \, \text{(} \textsc{countdown} \, \wedge \, \textsc{zero} \text{)} \, \vee \, \textsc{anykey} \\
             && \quad \ \ \leftrightarrow \LTLnext \, \upd{\name{beep}}{\name{true}} \, \big) \\[0.2em]

    \psi_{9} & := & \upd{\name{screen}}{\name{display}~\name{time}}

  \end{array}
\end{equation*}
\begin{figure}
\begin{lstlisting}[style=tsl,basicstyle=\tiny]
COUNTUP = [ time <- countup time dt ];
COUNTDOWN = [ time <- countdown time dt ];
INCMIN = [ time <- incMinutes time ];
INCSEC = [ time <- incSeconds time ];
IDLE = [ time <- time ];
ZERO = eq time zero();
RESET = btnMin && btnSec;
COUNTING = COUNTUP || COUNTDOWN;
ANYKEY = press btnMin || press btnSec
  || press btnStartStop;
START = press btnStartStop && !press btnMin
  && !press btnSec;
STARTANDMIN = press btnStartStop && press btnMin
  && X !btnSec && X X !btnSec;
STARTANDSEC = press btnStartStop && press btnSec
  && X !btnMin && X X !btnMin;

xor x y = !(x <-> y);
press x = !x && X x;
tillAnyInput x = (x && !ANYKEY) W (RESET || x && ANYKEY);

initially guarantee {
 !COUNTING && (X COUNTING -> START);
 !INCSEC && !INCMIN;
 [ beep <- False() ];
}

always guarantee {
 RESET <-> [ time <- zero() ];
 !COUNTING && press btnMin && X !btnSec <-> X INCMIN;
 !COUNTING && press btnSec && X !btnMin <-> X INCSEC;
 ZERO -> ((IDLE && START -> X tillAnyInput COUNTUP)
   W (INCMIN || INCSEC));
 INCMIN || INCSEC -> ((!COUNTING && START ->
   X tillAnyInput COUNTDOWN) W X ZERO);
 COUNTING && ANYKEY && X !RESET -> X tillAnyInput IDLE; 
 !COUNTING && (STARTANDMIN || STARTANDSEC)
   -> X X tillAnyInput COUNTDOWN;
 xor [ beep <- True() ] [ beep <- False() ];
 X (COUNTDOWN && ZERO) || ANYKEY <->  X [ beep <- True() ];
 [ dsp <- display time ]; 
}
\end{lstlisting}
\caption{The full timer specification, given in the used plain text specification
  format.}
\label{fig:codetsl}
\end{figure}
\!\!The function~$ \textit{tillAnyInput} $ is a helper function to denote that a
condition must be satisfied until either the system is reset or
any button gets pressed. It is defined as:
\begin{equation*}
  \begin{array}{rcl}
    \textit{tillAnyInput}~x & := & \text{(}x \,\wedge\, \neg \textsc{anykey} \text{)}
                                   \\
    && \quad \ \LTLweakuntil \; \text{(}\textsc{reset} \, \vee \, x \,\wedge\, \textsc{anykey} \text{)}
  \end{array}
\end{equation*}
The final specification is given by
$ \varphi = \psi_{\textit{init}} \wedge \LTLglobally
\ensuremath{\scalebox{1.3}{\ensuremath{\wedge}}}_{i = 1}^{9} \psi_{i}
$, where $ \psi_{\textit{init}} $ adds some remaining initial
conditions. The full specification, using our plain text specification
format, is also given in \cref{fig:codetsl}. Note that the various
$ \LTLnext $ operations in the formulas~$ \psi_{i} $ are always
necessary, since the ``being pressed'' condition requires a change of
the input, which is only observable by comparing the currently provided value
with the previous one.

For the development of such specifications, the designer also gets
feedback from the synthesis engine. For example, the condition
of $ \psi_{2} $ requires \name{btnSec} to not being pressed in order
to increase the minutes of the counter. Without this condition, the
synthesis engine would return an unrealizabilty result, since
increasing minutes would conflict with setting the time to zero on
a potential reset, which also requires \name{btnMin} to be pressed.

After the development of the specification is finished, the synthesis
automatically creates a CFM that satisfies the specified control
behavior. In the next step the CFM then is specialized towards the
specific application context.

\section{Code Generation}
\label{sec:synth}

We present a system for the generation of FRP program code from the output of $ \TSL $ synthesis.
The user initially provides a CFM that was synthesized from a $ \TSL $ specification over a set of predicate and function terms.
The user specifies a target FRP abstraction, and receives an executable Haskell FRP program in the library of their choosing.

Our approach takes a multi-stage approach, whereby the TSL specification is first used to generate a control flow model (CFM).
The CFM is an abstract representation of the temporal changes that the FRP program must implement in order to satisfy the TSL specification.
In particular, a CFM maps the input signals through various function and predicate terms to the output signals.
We only consider valid CFMs, where for every cycle created by mapping an output back to an input, there is at least one \textit{cell}.
A cell is a memory unit with delay, so that at one time step a value may be stored, and at the next that value is retrieved.
The concept of a cell is analogous to ArrowLoop~\cite{PatersonRA:notation}, or a register in hardware.

The synthesis from TSL to CFM is the most computationally expensive and may result in an unrealizable result, in which case synthesis terminates with no solution.
We omit a detailed description of the generation of the CFM from a TSL specification, and instead direct the reader to related work on this topic~\cite{tsl}.
Here we focus on how the generality of the CFM is utilized to generate framework-independent FRP code.
If a satisfying CFM is found during the first stage, a user specifies a target FRP abstraction (Applicative, Monad, Arrow) that should be used to generate the FRP program code from the CFM.

To clarify the translation process from a CFM to FRP code, we give a formal definition of a CFM.
A CFM~$ \cfm $ is a tuple $ \cfm = (\inames, \onames, \cells, \vertices, \labeling, \dependencies), $ 
where 
  $ \inames $ is a finite set of inputs,
  $ \onames $ is a finite set of outputs, 
  $ \cells $ is a finite set of cells,
  $ \vertices $ is a finite set of vertices,
  $ \labeling \from \vertices \to (\fnames \cup \logic \cup \mutex) $ assigns a vertex a signal function
     (either a function~$ \name{f} \in \fnames $, a predicate~$ \name{p} \in \pnames $, a logic operator in \logic lifted to the signal level, or a mutex selector \mutex lifted to the signal level).
The set of logic operators are the standard Boolean operators (and, or, not), and the mutex selectors are signal functions pattern matching on one input signal to select among the other input signals for output.
Finally, a CFM also contains a dependency relation 
\begin{equation*}
  \dependencies \from (\onames \cup \cells \cup \vertices) \times
  \nats \to (\inames \cup \cells \cup \vertices \cup \set{ \bot })
\end{equation*}
relating every output, cell, and vertex to a set of inputs, cells, or vertices. 
The dependency relations defines the wiring between signal functions.
The selector $ n \in \nats $ argument allows us to specify a specific connection, since a signal function may have multiple inputs.
Outputs and cells~$ \name{s} \in \onames \cup \cells $ always have only a single input signal stream, so the first selector has some non-bottom value ($ \delta(s, 0) \not\equiv \bot $) and any larger selector is undefined (\mbox{$ \forall m > 0 .\ \delta(\name{s}, m) \equiv \bot $}).
In contrast, for vertices~$ x \in \vertices $ the number of input signals $ n \in \nats $ match the arity of the assigned function or predicate $ \labeling(x) $.
This means $ \forall m \in \nats .\ \delta(x, m) \equiv \bot \leftrightarrow m > n $.
We only consider valid CFMs, where a CFM is valid if it does not contain circular dependencies,
i.e., on every cycle induced by $ \delta $ there must lie at least a single cell.
As an example, a CFM would contain a circular dependency if given $x, y \in \vertices$, $\dependencies(x,0) = y$ and $\dependencies(y,0) = x$.
Such a CFM, if rendered as an Arrow function, would enter an infinite loop, and in the best case, generate the runtime error \texttt{<<loop>>}.

Given a CFM that satisfies the \TSL specification, we next convert it into a template for our FRP program.
The CFM is transformed via a syntactic transformation into an FRP program in the abstraction of the user's choice, as a function that is parameterized over the named function and predicate terms, as shown in Fig.~\ref{fig:controlTemplate}. 
The user then provides implementations of the function and predicate terms that complete the construction of the FRP program based on the generated template.

\begin{figure}

\begin{subfigure}{\columnwidth}
\begin{lstlisting}
control
  :: _ signal  -- FRP abstraction
  => _         -- cell implementation
  -> (_ -> _)   -- functions and predicates
  -> _         -- initial values
  -> signal _  -- input signals
  -> signal _  -- output signal
\end{lstlisting}
\caption{The general template of the type signature for the control block of the synthesized FRP program}
\label{fig:controlTemplateGeneral}
\end{subfigure}

\begin{subfigure}{\columnwidth}
\begin{lstlisting}
control
  :: Applicative signal
  => (forall p. p -> signal p -> signal p)
  -> (a -> Bool)  -- press
  -> (b -> c)     -- display
  -> (b -> b)     -- increment
  -> b           -- initial value: count
  -> c           -- initial value: screen
  -> signal a    -- button (input)
  -> ( signal b  -- count (output)
    , signal c  -- screen (output)
    )
\end{lstlisting}
\caption{An example instantiation of the type signature for the control block of the button (as described in the introduction) as it has been specialized for Applicative FRP.}
\label{fig:controlTemplateTimer}
\end{subfigure}

\caption{The control block follows a general type signature template across FRP abstractions.}
\label{fig:controlTemplate}
\end{figure}

The CFM transformation is modularized to fit any FRP library that is at least as powerful as CCA~\cite{ploeg2015frpnow,perez2016yampa,murphy2016livefrp,patai2010efficient,FRPzoo}.
The key insight is that first-order control along with a loop describes the expressive power of both CCA and the CFM model generated from the techniques of~\cite{tsl}.
Although many FRP libraries support more powerful operations than CCA, \eg \texttt{switch} in \texttt{Yampa}, we do not need to utilize these in the synthesis procedure, and thus can generalize synthesis to target any FRP library that is at least as expressive as CCA.

Recall that in $\TSL$, output signals can be written at the current time~$t$, and be read from at time $t+1$.
To implement this in the FRP program, we use the concept of a cell in the CFM.
In the translation, we allow a space for the user to provide an implementation of the cell that is specialized to their FRP library of choice.
In the case of CCA, this is the \verb|loopD| combinator. 
The \verb|loopD| combinator pipes the output values back to the input to allow them to be read at time $t+1$.  
Since a system may require output values at time $t=0$, the user must also provide initial values to $\onames$.  

\begin{figure}
\begin{lstlisting}[mathescape=true]
control =  
  rec
-- gather values from the previous time step
    $\forall c \in \cells.\ c \leftarrow \dependencies(c,0)$
-- gather applications of $\fnames$ and $\pnames$
    $\forall v \in (\vertices \cap (\fnames \cup \pnames)).\ v \leftarrow $($\dependencies(v,0), ..., \dependencies(v,n)$)
-- compute control signals from $c$s, $v$s, and $ctrl$s
    $\forall ctrl \in (\vertices \cap \logic).\ ctrl \leftarrow $($\dependencies(ctrl,0), ..., \dependencies(ctrl,n)$)
-- use control signals to select from $\fnames$ and $\pnames$ applications
    $\forall m \in (\vertices \cap \mutex).\ m \leftarrow $($\dependencies(m,0), ..., \dependencies(m,n)$)

-- output signals take the signals from either the $c$s, $v$s, or $m$s as specified by $\dependencies$
  return $\forall o \in \onames.\ o \leftarrow \dependencies(o,0)$

\end{lstlisting}
\caption{The general code template for the control block of the synthesized FRP program. The exact syntax for \texttt{rec}, assignment, and outputting signals varies across different FRP abstractions.}
\label{fig:controlTemplateCode}
\end{figure}

\subsection{Properties of Synthesis}
\label{sec:synthProp}

For a full description of the formal properties of TSL synthesis, see the work of~\cite{tsl}.
In summary the synthesis procedure is sound, but not complete.
From a programming languages design perspective, this means that ``compilation'' (synthesis) of a specification may not terminate, but when it does terminate, it will generate code that satisfies the specification.

In the translation of the CFM, we use Casual Commutative Arrows (CCA) as the target conceptual model for the transformation.
One interesting note about this is that CCA does not allow the \texttt{arrowApply} function.
The \texttt{arrowApply} (also called \texttt{switch}) function is a higher-order arrow that allows for dynamically replacing an arrow with a new one that arrives on an input wire, enforcing a static structure on the generated program.
As one of our target domains is hardware (using \texttt{ClaSH}), this would cause problems, as self-reconfiguring hardware is out-of-scope for this work.
The insight provided by prior work on CCA~\cite{jfp/LiuCH11} found that, in general, the expressive power of higher-order arrows makes automatic optimization more difficult.
Furthermore, for most FRP programs, first-order switch is more than enough~\cite{Winograd-Cort:2014}.

With respect to the synthesis procedure, this is a fundamental restriction related to the specification logic~\TSL.
With \TSL, every update term \upd{x}{y} is lifted to an arrow that updates x with y over time.
Since in \TSL updates are fixed by the specification, so too must the arrow structure be fixed in synthesis.

Note that having a fixed arrow structure disallows higher-order arrows, but higher-order functions can still be passed along the wires.
As an example, we may have a function term \mbox{\codeinline{app::(a->b)->a-> b}}
and signals \codeinline{f::a->b} and \codeinline{x::a}.
A simple specification making use of higher order functions then could state that the system should always apply the incoming higher-order function to the incoming value: $\LTLglobally \, \upd{\name{x}}{\name{app}(\name{f},\name{x})}$.
Recall that, absent any user provided assumptions, the synthesis procedure allows the value of all symbols to change at every time step.
Proper reasoning over higher-order functions then requires added assumptions in practice, depending on the specific use case.

Additionally, a key difference between arrows and circuits is that arrows are able to carry state that tracks the application of each arrow block.
By using CCA, a user may write \TSL specifications about stateful arrows that are still handled correctly by the synthesis procedure.
To this end, we only synthesize programs that obey the \textit{commutativity} law~\cite{jfp/LiuCH11,yallop2016causal}, restated below, 
which ensures that arrows cannot carry state that influences the result of composed computations.
\begin{equation*}
first\ f\ \arrComp second\ g\ =\ second\  g\ \arrComp first\ f
\end{equation*}
Imagine an arrow with a global counter to track data of a buffer. Since addition is commutative, this arrow respects the commutativity law.
However, non-commutative state is possible as well.
For example, when building GUIs with Arrowized FRP~\cite{UISF}, the position of each new UI element depends on the order of the previously laid out elements.
Due to the commutativity of the Boolean operators, the commutativity of CCA is a necessary precondition for synthesis of a \TSL specification.
Specifically, the commutativity of logical conjunction allows the solution to update signals in any arbitrary order.
Thus, the correctness of the \TSL synthesis relies on commutativity of composition, which is naturally modeled with CCA's commutativity law.

\subsection{Example: Kitchen Timer}

\begin{figure}

\begin{subfigure}{\columnwidth}
\begin{lstlisting}
control
  :: (MonadFix monad, Applicative signal)
  => (forall p. p -> signal p -> monad (signal p))
\end{lstlisting}
\vspace{-0.5em}
\begin{center}
  $ \cdots $
\end{center}
\vspace{-0.5em}
\begin{lstlisting}
  -> signal a -> monad (signal b, signal c)
\end{lstlisting}
\caption{The Monadic control for the button from the introduction.}
\label{fig:monadicTimer}
\end{subfigure}

\begin{subfigure}{\columnwidth}
\begin{lstlisting}
control
  :: (Arrow signal, ArrowLoop signalfunction)
  => (forall p. p -> signalfunction p p))
\end{lstlisting}
\vspace{-0.5em}
\begin{center}
  $ \cdots $
\end{center}
\vspace{-0.5em}
\begin{lstlisting}
  -> signalfunction a (b, c)
\end{lstlisting}
\caption{The Arrowized control for the button from the introduction.}
\label{fig:arrowTimer}
\end{subfigure}

\begin{subfigure}{\columnwidth}
\begin{lstlisting}
control
  :: HiddenClockReset domain gated synchronous
\end{lstlisting}
\vspace{-0.5em}
\begin{center}
  $ \cdots $
\end{center}
\vspace{-0.5em}
\begin{lstlisting}
  -> Signal domain a
  -> (Signal domain b, Signal domain c)
\end{lstlisting}
\caption{The Applicative (specialized to \texttt{ClaSH}) control for the button from the introduction.}
\label{fig:clashControl}
\end{subfigure}

\caption{The abbreviated type signatures for the various synthesized control blocks for each FRP framework abstraction.}
\label{fig:timerControl}

\end{figure}

\begin{figure}
  \centering
  \includegraphics[scale=0.061]{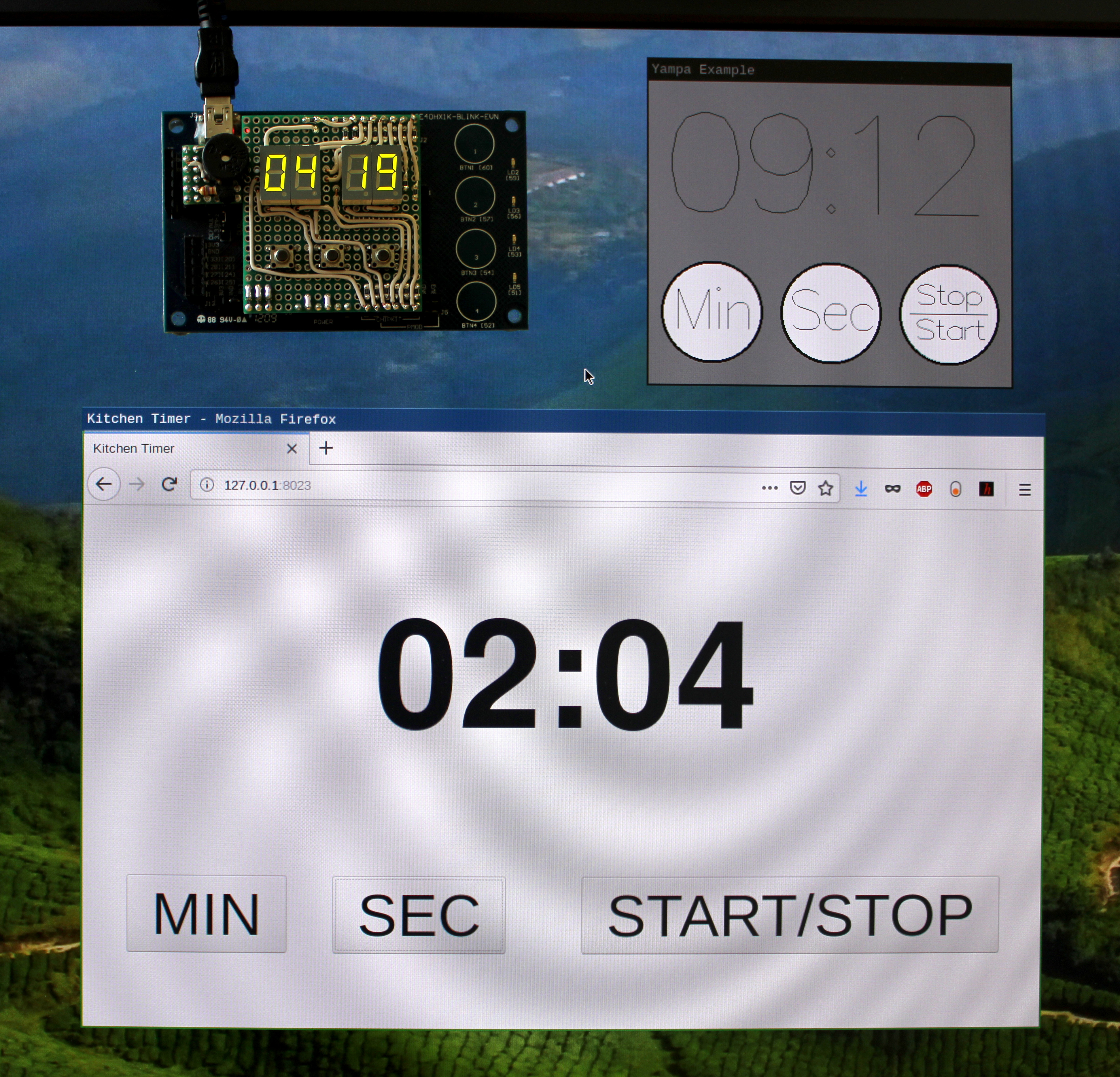}
  \caption{Timer applications: the hardware application
    built with \texttt{ClaSH} is on the top left. The top right shows
    the desktop application built using \texttt{Yampa} and the at
    bottom the web application built using
    \texttt{threepenny-gui}. All are synthesized from the same CFM.}
  \label{fig:apps}
\end{figure}

We revisit the Kitchen Timer application introduced in \cref{sec:motiv} to show the concrete code that is generated.
From the TSL specification, we first generate a CFM using our TSL synthesis toolchain together with the LTL synthesizer \name{strix}~\cite{DBLP:conf/cav/MeyerSL18} [version \texttt{18.04}]. The resulting CFM utilizes six additionally synthesized cells and consists of 1188 vertices.
This CFM is then transformed into a control structure for each of the different application domains.
In \cref{fig:timerControl}, we show how the template described in Sec.~\ref{sec:synth} is specialized to each of the three application domains: the desktop program is built with \texttt{Yampa} [version~\texttt{0.13}] and the web app with \texttt{threepenny-gui} [version~\texttt{0.8.3.0}]. Both have been built using stack\footnote{\url{https://www.haskellstack.org}} on \texttt{lts-13.17} [\texttt{ghc-8.6.4}]. For building the hardware implementation, we first use the functional hardware description language \texttt{ClaSH}\footnote{\url{https://github.com/clash-lang/clash-compiler}~[commit:~\texttt{fff4606}]} to generate verilog-code, which then is turned into the \texttt{blif} format using the open synthesis suite \texttt{yosys}\footnote{\url{https://github.com/YosysHQ/yosys}~[commit:~\texttt{70d0f38}]}. Afterwards, the generated \texttt{blif}-file is placed using the place-and-route tool \texttt{nextpnr}\footnote{\url{https://github.com/YosysHQ/nextpnr}~[commit:~\texttt{5344bc3}]}. The packaged result is then uploaded to an iCEblink40HX1K Evaluation Kit Board from Lattice Semiconductor, featuring an ICE40HX1K FPGA with 100 IO-pins and 1280 logic cells, additionally equipped with the required hardware components. The interfaces to the corresponding timer applications are depicted in \cref{fig:apps}. The synthesis and compilation times of the different tools are depicted in \cref{table:compileTimes}. A video demonstration of the three different timers can also be found at:
\begin{equation*}
  \text{\url{www.youtube.com/channel/UCLepxl4YhH1yryPJsTEf4eQ}}
\end{equation*}

\begin{table}[b]
\begin{tabular}{|l|c|}
  \hline
  \textbf{Process} &   \textbf{Time (sec)}  \\ \hline \hline
  Synthesis $ \rightarrow $ \texttt{strix} & 4.965  \\ 
  Compilation & \\
  \quad Desktop $ \rightarrow $ \texttt{Yampa} & 19.403  \\
  \quad Web $ \rightarrow $ \texttt{threepenny-gui} & 18.344  \\
  \quad Hardware & \\
  \qquad $ \rightarrow $ \texttt{ClaSH} & 11.218 \\
  \qquad $ \rightarrow $ \texttt{yosys} & 6.405 \\
  \qquad $ \rightarrow $ \texttt{nextpnr} & 7.276 \\ \hline
\end{tabular}
\caption{Synthesis and compilation times for creating the different
  timer applications from the TSL specification.}
\label{table:compileTimes}
\vspace{-1.4em}
\end{table}

\texttt{Yampa} uses Arrowized FRP which easily fits into the general interface for Arrows that we provide (Fig.~\ref{fig:arrowTimer}).
Likewise, \texttt{threepenny-gui} uses a Monadic FRP (where the signals themselves are Applicative) which also easily fits into our general interface for Monadic FRP (Fig.~\ref{fig:monadicTimer}).
Finally, for \texttt{ClaSH}, we use a mostly Applicative interface, that is specialized to handle the peculiarities of hardware (which needs explicit clocks, as opposed to more traditional FRP frameworks).
If we wanted to support other libraries with explicit clocks, for example, as presented in~\cite{DBLP:conf/haskell/Barenz018}, we would need a specialized module for this - although the customization is limited mostly to the type signature generation as shown in Fig~\ref{fig:clashControl}. 

Each control block requires the user to provide a cell implementation.
Both \texttt{Yampa} and \texttt{ClaSH} provide native implementations of the concepts, as shown in Fig~\ref{fig:cellImplementations}.
Although \texttt{threepenny-gui} does not provide the exact implementation of a cell, as we require in our synthesized control block, it can be easily implemented using the available primitives of the library.

\begin{figure}
\begin{lstlisting}
-- yampa
iPre :: SF a a

-- clash
register
  :: HiddenClockReset domain gated synchronous
  => a -> Signal domain a -> Signal domain a

-- reactivebanana / threepennygui
cell
  :: MonadMoment / MonadIO m
  => a -> Behavior a -> m (Behavior a)
cell v x = stepper v (x <@ allEvents)

-- reflex
cell
  :: Reflex t
  => b -> Behavior t b -> Behavior t b
cell v x =
  hold v (tag x allEvents)
\end{lstlisting}
\caption{The implementations for a cell in the CFM is commonly found across FRP libraries, or easliy re-implemented.}
\label{fig:cellImplementations}
\end{figure}

\section{Related Works}

There are various lines of work that are related to our approach.
While we draw inspiration from these research directions, each one, on

\subsection{Temporal Types for FRP}

FRP is a programming paradigm for computations over time, and, hence, a natural extension is to investigate type systems to be able to reason about time.
A correspondence between LTL and FRP in a dependently typed language was discovered simultaneously by~\cite{plpv/Jeffrey12,jeltsch2012towards}.
In this formulation, FRP programs are proofs and LTL propositions are reactive, time-varying types that describe temporal properties of these programs.
In establishing the connection between logic and FRP, these LTL types are also used to ensure causality and loop-freeness on the type level.

Dependent LTL types are a useful extension to FRP that provides insight into the underlying model of FRP, but does not lend itself to control flow synthesis.
In the work of Jeffery and Jeltsch, the types describe the input/output change over time for each arrow.
Using these LTL types, only arrows adhering to sensible temporal orderings (\eg computations only depend on past values) will be well typed.
However, as with any other FRP system, the temporal control flow of function applications in the program is fixed by the code.
A similar approach was used by~\cite{krishnaswami2013higher} to make a temporal type system that ensures there are no spacetime leaks in an well typed FRP program.
Work on reasoning about FRP using temporal logics also includes~\cite{sculthorpe2010keeping}, although this setting considered dynamic network structure, as allowed by higher-order arrows. 
Further properties beyond safety have been expressed in FRP, for instance using a type extension to enforce fairness properties~\cite{Cave2014Fair}.

In contrast, we use the logic \TSL, for a fine-grained description of function application behavior which cannot be expressed within pure LTL. 
The synthesis procedure of TSL determines a temporal control flow of functions, where the \TSL specifications determines the transformations to be applied at each point in time.
In addition to the logical specifications, the synthesis is also constrained by the types of functions appearing in the specification.
Since the types of all functions are fixed for every time step, the type system can be lifted to the specification.
If the specification is well typed, synthesis is guaranteed to yield a well typed program.

One connection to our work however is the implications of the fact that the Curry-Howard correspondence extends to FRP and LTL.
In the aforementioned work, LTL propositions are types for FRP programs.
If a proof of a TSL proposition can be interpreted as a program, one might expect that there is some corresponding type system to \TSL.
We leave such explorations to future work.

\subsection{Synthesis of Reactive Programs}

A distinguishing feature of our approach is the connection to an actual programming paradigm, namely FRP.
Most reactive synthesis methods instead target transition systems or related formalisms such as finite state machines.
The idea to synthesize programs rather than transition systems was introduced in~\cite{madhusudan2011synthesizing}. 
In his work, an automaton is constructed that works on the syntax tree of the program, which
  makes it possible to obtain concise representations of the implementations, and to determine how many program variables are needed to realize a particular specification.
Unlike our FRP programs, Madhusudan's programs only support variables on a finite range of instances.

Another related approach is the synthesis of synchronization
primitives introduced in~\cite{bloem2012synthesis} for the purpose of allowing sequential programs to be executed in
parallel. Similar to our synthesis approach,
uninterpreted functions are used to abstract from implementation
details.  However, both the specification mechanism (the existing
program itself is used as the specification) and the type of programs
considered are completely different from TSL and FRP.

\subsection{Logics for Reactive Programs}

Many logics have been proposed to specify properties of reactive programs.
Synthesis from Signal Temporal Logic~\cite{hscc/sanjit15} focuses on modeling physical phenomena on the value level, introducing continuous time to the model and resolving to a system of equations.
The approach allows for some different notion of data embedded into the equations.
While more focused on the data level, the handling on continuous time might provide inspiration for future extensions to \TSL to more explicitly handle continuous time. 
    
Another logic that has been proposed, Ground Temporal Logic~\cite{cyrluk1994ground}, 
is a fragment of First Order Logic equipped with temporal operators, where it is not allowed to use quantification.
Satisfiability and validity problems are studied, with the result that only a fragment is decidable.
However, specifications expressed in Ground Temporal Logic, as well as their motivations, are completely different from our goals.

\subsection{Reasoning-based Program Synthesis}

Reasoning-based synthesis~\cite{osera2015type,solarLezama13, 
kuncak2010complete, vechevYY13} is a major line of work that has been 
mostly, but not entirely, orthogonal to reactive synthesis. 
While reactive synthesis has focused on the complex control aspects of reactive systems, 
  deductive and inductive synthesis has been concerned with the data transformation aspects in non-reactive, sequential programs. 
Our work is most related to Sketching~\cite{solarLezama13}, since in Sketching the user provides the control structure and synthesizes the transformations while with TSL we synthesize the control structure and leave the transformations to the user.

The advantage of deductive synthesis is that it can handle systems with 
complex data. Its limitation is that it cannot handle the continuous interaction between 
the system and its environment, which is typical for many applications, such as for 
cyber-physical systems. This type of interaction can be handled by reactive synthesis, 
which is, however, typically limited to finite states and can therefore not be used in 
applications with complex data types. Abstraction-based approaches can be seen as a link between deductive and reactive synthesis~\cite{Dimitrova2012,Beyene:2014:CAS:2535838.2535860}.

Along the lines of standard reactive synthesis, our work is focused on synthesizing control structures. We extend the classic approach by also allowing the user to separately provide implementations of data transformations.
This is useful in the case where the value manipulations are unknown or beyond the capability of the synthesis tool.
For example, a user may want to synthesize an FRP program that uses closed source libraries, which may not be amenable to deductive synthesis.
In this case, the user can only specify that certain functions from that API should be called under certain conditions, but cannot and may not want to reason about their output.

\section{Conclusions}

In this work we have presented a detailed account of how to transform Control Flow Models into framework-independent FRP code. With this transformation, we utilize TSL synthesis as presented in~\cite{tsl} to build a complete toolchain for synthesizing Functional Reactive Programs.

So far we have used a discrete time model in our formalization, however, the behavior of the kitchen timer is in fact sampling rate independent (Continuous Time FRP). Sampling rate independence is guaranteed in TSL as long as the next operator is not used. However, the relation between TSL with the next operator and Continuous Time FRP still needs to be explored.

In another direction, the usual way  in FRP to distinguish between continuous and discrete behaviors is to use signals and events. So far we have embedded all data into signals. It is open to future work how to utilize events natively. Future directions for improvements to usability include integrating FRP synthesis more tightly into the typical programming workflow - for instance by allowing TSL specification to be used inline in code with QuasiQuoters~\cite{mainland2007s}.

\bibliography{biblio}

\end{document}